\documentclass[a4paper]{lipics-v2021}
\usepackage{cite}
\usepackage{amsmath,amssymb,amsfonts}
\usepackage{algorithmic}
\usepackage{graphicx}
\usepackage{textcomp}
\usepackage{xcolor}
    
\usepackage{footmisc}

\usepackage{todonotes}
\usepackage{color}
\usepackage{xcolor}

\usepackage{xcolor}
\usepackage[ruled,vlined, linesnumbered]{algorithm2e}

\usepackage{siunitx}
\usepackage[nolist]{acronym}
\usepackage{subcaption}

\usepackage{hyperref} 
\usepackage{mathptmx}
\DeclareMathAlphabet{\mathcal}{OMS}{cmsy}{m}{n}

\usepackage[shortlabels]{enumitem}
\setlist[enumerate]{nosep}

\usepackage{etoolbox}\AtBeginEnvironment{algorithmic}{\small}
\usepackage{balance}

\newacro{GCL}[GCL]{Gate-Control List}
\newacro{TSN}[TSN]{Time-Sensitive Networking}
\newacro{TAS}[TAS]{Time-Aware Shaper}
\newacro{TESLA}[TESLA]{Timed Efficient Stream Loss-Tolerant Authentication}
\newacro{CPS}[CPS]{Cyber-Physical System}
\newacro{CP}[CP]{Constraint Programming}
\newacro{SA}[SA]{Simluated Annealing}
\newacro{MILP}[MILP]{Mixed Integer Linear Programming}
\newacro{MTU}[MTU]{Maximimum Transmission Unit}
\newacro{WCET}[WCET]{Worst-case Execution Time}

\title{Dependability-Aware Routing and Scheduling for Time-Sensitive Networking}
\author{Niklas Reusch}{Technical University of Denmark Kongens Lyngby, Denmark}{nikre@dtu.dk}{}{}
\author{Silviu S. Craciunas}{TTTech Computertechnik AG, Vienna, Austria}{silviu.craciunas@tttech.com}{}{}
\author{Paul Pop}{Technical University of Denmark Kongens Lyngby, Denmark}{paupo@dtu.dk}{}{}

\authorrunning{N. Reusch et. al.} 

\Copyright{Mohammadreza Barzegaran, Niklas Reusch, Luxi Zhao, Silviu S. Craciunas, and Paul Pop} 



\category{} 

\relatedversion{} 
\hideLIPIcs  

\EventEditors{}
\EventNoEds{2}
\EventLongTitle{}
\EventShortTitle{}
\EventAcronym{}
\EventYear{}
\EventDate{}
\EventLocation{}
\EventLogo{}
\SeriesVolume{}
\ArticleNo{}

\begin{document}
\nolinenumbers

\maketitle

\begin{abstract}
Time-Sensitive Networking (TSN) extends IEEE 802.1 Ethernet for safety-critical and real-time applications in several areas, e.g., automotive, aerospace or industrial automation. However, many of these systems also have stringent security requirements, and security attacks may impair safety. Given a TSN-based distributed architecture, a set of applications with tasks and messages, as well as a set of security and redundancy requirements, we are interested to synthesize a system configuration such that the real-time, safety and security requirements are upheld. We use the Timed Efficient Stream Loss-Tolerant Authentication (TESLA) low-resource multicast authentication protocol to guarantee the security requirements, and redundant disjunct message routes to tolerate link failures. We consider that tasks are dispatched using a static cyclic schedule table and that the messages use the time-sensitive traffic class in TSN, which relies on schedule tables (called Gate Control Lists, GCLs) in the network switches. A configuration consists of the schedule tables for tasks as well as the disjoint routes and GCLs for messages. We propose a Constraint Programming-based formulation which can be used to find an optimal solution with respect to our cost function. Additionally, we propose a Simulated Annealing based metaheuristic, which can find good solution for large test cases. We evaluate both approaches on several test cases.
\end{abstract}

\section{Introduction}
\begin{figure}[t]
    \centering
	\includegraphics[width=0.65\textwidth]{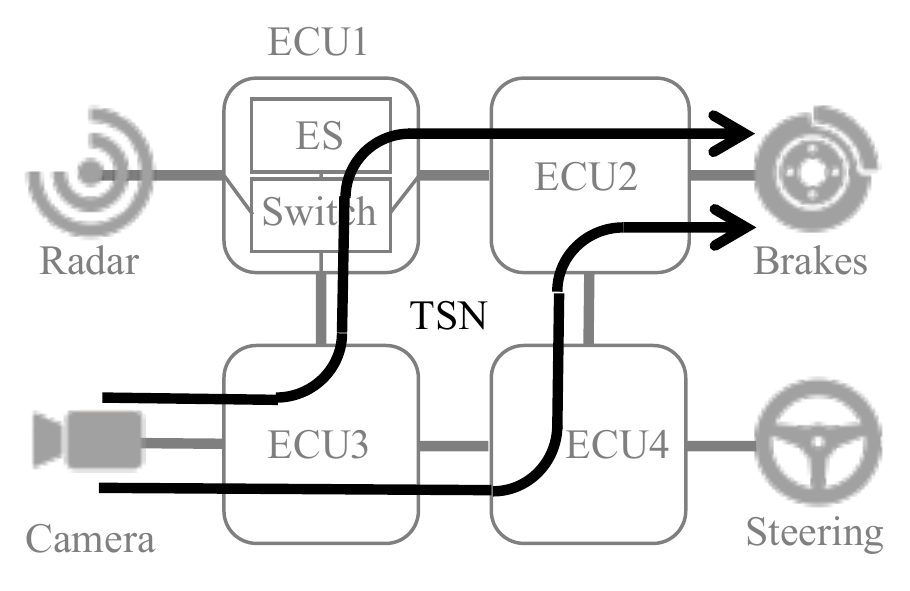}
	\caption{Example automotive TSN-based CPS with redundant routing}
	\label{fig:auto}
\end{figure}

Many modern safety-critical real-time systems are implemented on distributed architectures. They integrate functions with different security and safety requirements over the same deterministic communication network. For example, the network in a modern vehicle has to integrate high-bandwidth video and LIDAR data for Advanced Driver Assistance Systems (ADAS) functions with the highly critical but low bandwidth traffic of e.g. the powertrain functions, but also with the best-effort messages of the low-criticality diagnostic services. See \autoref{fig:auto} for an example network architecture of a modern vehicle.

\ac{TSN}~\cite{8021tsn}, which is becoming the standard for communication in several application areas, e.g. automotive to industrial control, is comprised of a set of amendments and additions to the IEEE~802.1 standard, equipping Ethernet with the capabilities to handle real-time mixed-criticality traffic with high bandwidth. A TSN network consists of several end-systems, that run mixed-criticality applications, interconnected via network switches and physical links. Available traffic types are Time-Triggered (TT) traffic for real-time applications, Audio-Video Bridging (AVB) for communication that requires less stringent bounded latency guarantees, and Best-Effort (BE) traffic for non-critical traffic.

We assume that safety-critical applications are scheduled using static cyclic scheduling and use the TT traffic type with a given \textit{Redundancy Level} (RL) for communication. We consider that the task-level redundancy is addressed using solutions such as replication \cite{replication}, and we instead focus on the safety and security of the communication in \ac{TSN}. The real-time safety requirements of critical traffic in \ac{TSN} networks are enforced through offline-computed schedule tables, called Gate Control Lists (GCLs), that specify the sending and forwarding times of all critical frames in the network. Scheduling time-sensitive traffic in \ac{TSN} is non-trivial (and fundamentally different from e.g. TTEthernet), because \ac{TSN} does not schedule communication at the level of individual frames as is the case in TTEthernet. Instead, the static schedule tables (GCLs) governs the behavior of entire traffic classes (queues) which may lead to non-deterministic frame transmissions ~\cite{craciunas_rtns_16}.

Since link and connector failures in \ac{TSN} could result in fatal consequences, the network topology uses redundancy, e.g., derived with methods such as~\cite{VoicaBahramRedundancy}. In \ac{TSN}, IEEE~802.1CB Frame Replication and Elimination for Reliability (FRER) enables the transmission of duplicate frames over different (disjoint) routes, implementing merging of frames and discarding of duplicates.

Nowadays modern \acp{CPS} are becoming increasingly more interconnected with the outside world opening new attack vectors~\cite{industry4.0_security, automotive_security_survey} that may also compromise safety. Therefore, the security aspects should be equally important to the safety aspects. \ac{TESLA}~\cite{tesla} has been investigated as a low resource authentication protocol for several networks, such as FlexRay and TTEthernet~\cite{security_aware_tte} networks. Adding security mechanisms such as \ac{TESLA} after the scheduling stage is oftentimes not possible without breaking real-time constraints, e.g. on end-to-end latency, and degrading the performance of the system~\cite{security_aware_tte}. Thus we consider \ac{TESLA} and the overhead and constraints it imposes part of our configuration synthesis problem formulation.

\subsection{Related Work}
Scheduling for TSN networks is a well-researched problem. It has been solved for a variety of different traffic type combinations (TT, AVB, BE) and device capabilities using methods such as Integer Linear Programming (ILP), Satisfiability Modulo Theories (SMT) or various metaheuristics such as tabu search~\cite{craciunas_rtns_16, SernaRTAS18, Frank16, AVBAwareRoutingScheduling}.

Routing has also been extensively researched~\cite{MulticastRoutingQoS, PacketRoutingSurvey}. The authors in \cite{AFDXRouting} presented an ILP solution to solve the routing problem for safety-critical AFDX networks. In \cite{TamasTTERouting} the authors used a tabu search metaheuristic to solve the combined routing and scheduling problem for TT traffic in TTEthernet. In \cite{QuorumcastRouting} the authors provide a simple set of constraints to solve a general multicast routing problem using constraint programming, which \cite{VoicaBahramRedundancy} builds on that to solve a combined topology and route synthesis problem. In \cite{LoadBalancingRouting} the authors use a load-balancing heuristic to distribute the bandwidth usage over the network and achieve smaller latency for critical traffic.

Multiple authors have also looked at the combined routing and scheduling problem. The authors in \cite{SuneRoutingScheduling} and \cite{QBVRouting} showed that they are able to significantly reduce the latency by solving the combined problem with an ILP formulation. In \cite{HeuristicRoutingScheduling} the authors presented a heuristic for a more complex application model that allows multicast streams. They were able to solve problems that were infeasible to solve using ILP or separate routing and scheduling.

Recently authors have started to present security- and redundancy-aware problem formulations. The authors in \cite{security_aware_tte} provided a security-aware scheduling formulation for TTEthernet using TESLA for authentication. In \cite{SecurityAwareRoutingAndScheduling} the authors solve the combined routing and scheduling problem and considered authentication using block ciphers. The authors in \cite{ReliabilityAwareRorutingScheduling} and \cite{TSNRoutingScheduling}, on the other hand, present a routing and scheduling formulation that is redundancy-aware but has no security considerations.

To the best of our knowledge our work is the first one to provide a formulation that is both security and redundancy-aware.

\subsection{Contributions}
In this paper, we address TSN-based distributed safety-critical systems and solve the problem of configuration synthesis such that both safety and security aspects are considered. Determining an optimized configuration means deciding on the schedule tables for tasks as well as the disjoint routes and GCLs for messages. Our contributions are the following:
\begin{enumerate}
    \item We apply \ac{TESLA} to \ac{TSN} networks considering both the timing constraints imposed by \ac{TSN} and the security constraints imposed by \ac{TESLA}.
    
    \item We formulate an optimization problem to determine: (i) the redundant routing of all messages; (ii) the schedule of all messages, encapsulated into Ethernet frames, represented by the GCLs in the network devices, and (iii) the schedule of all related tasks on end-systems.
    
    \item We extend our Constraint Programming (CP) formulation from \cite{RTSS_WIP} and propose a new Simulated Annealing (SA)-based metaheuristic to tackle large scale networks that cannot be solved with CP
    
    \item We evaluate the impact of adding the security from \ac{TESLA} on the schedulability of applications and we evaluate the solution quality and scalability of the \ac{CP} and \ac{SA} optimization approaches
\end{enumerate}

We introduce the fundamental concepts of \ac{TSN} in \autoref{sec:TSN} and of \ac{TESLA} in \autoref{sec:TESLA}.
In \autoref{sec:Models} we present the model of our system, consisting of the architecture of the network, applications running on this architecture. Additionally we present a threat model and how it is addressed by \ac{TESLA} with a security model. In \autoref{sec:Problem_Formulation} we formulate the problem we are solving using the established models and present an example. In \autoref{sec:Constraint_Formulation} and \autoref{sec:metaheuristic} we present the two different optimization approaches, CP and SA. Then, we evaluate these approaches using several test cases in \autoref{sec:Evaluation}. \autoref{sec:Conclusion} concludes the paper.

\section{Time-Sensitive Networking}
\label{sec:TSN}

\begin{figure}[!t]
    \centering
	\includegraphics[width=0.75\textwidth]{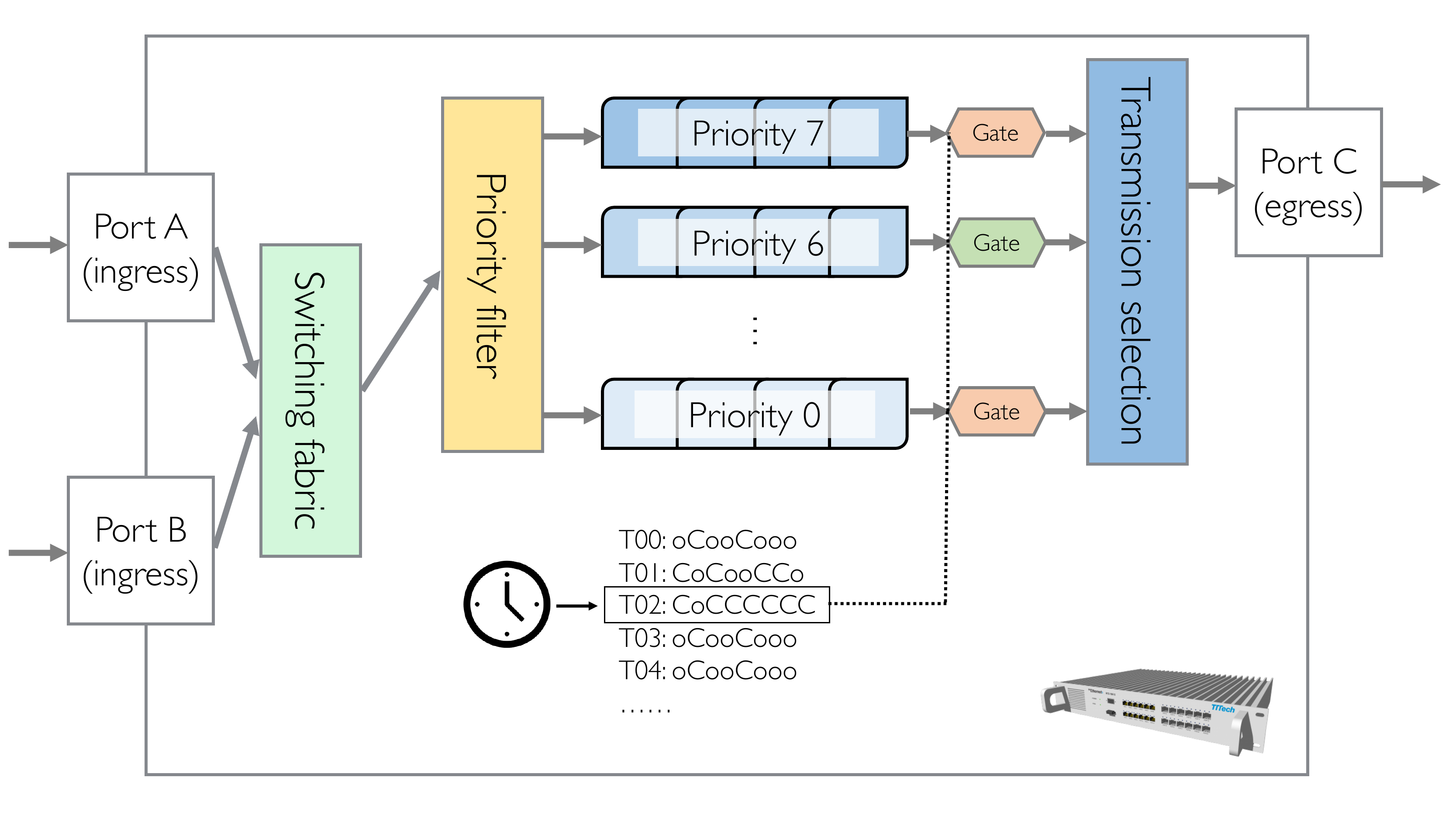}
	\caption{\label{fig:TSNswitch} Simplified TSN switch representation}
\end{figure}

Time-Sensitive Networking~\cite{8021tsn} has arisen out of the need to have more stringent real-time communication capabilities within standard Ethernet networks. Other technologies that offer real-time guarantees for distributed systems are TTEthernet (SAE AS6802~\cite{sae-as6802, steiner11:CRC}), PROFINET, and EtherCAT~\cite{4638425}. TSN comprises a set of (sub-)standards and amendments for the IEEE 802.1Q standard, introducing several new mechanisms for Ethernet bridges, extensions to the IEEE 802.3 media access control (MAC) layer, as well as other standards and protocols (e.g., 802.1ASrev). 

The fundamental mechanisms that enable deterministic temporal behavior over Ethernet are, on the one hand, the clock synchronization protocol defined in IEEE 802.1ASrev \cite{8021asrev}, which provides a common clock reference with bounded deviation for all nodes in the network, and on the other hand, the timed-gate functionality (IEEE 802.1Qbv~\cite{8021qbv}) enhancing the transmission selection on egress ports. The timed-gate functionality (IEEE 802.1Qbv~\cite{8021qbv}) enables the predictable transmission of communication streams according to the predefined times encoded in so-called Gate-Control Lists (GCL). A stream in TSN definition is a communication carrying a certain payload size from a talker (sender) to one or more listeners (receivers), which may or may not have timing requirements. In the case of critical streams, the communication has a defined period and a maximum allowed end-to-end latency.

Other amendments within TSN (c.f.~\cite{8021tsn}) provide additional mechanisms that can be used either in conjunction with 802.1Qbv or stand-alone. IEEE 802.1CB \cite{802.1CB} enables stream identification, based on e.g., the destination MAC and VLAN-tag fields in the frame, as well as frame replication and elimination for redundant transmission. IEEE 802.1Qbu \cite{8021qbu} enables preemption modes for mixed-criticality traffic, allowing express frames to preempt lower-priority traffic. IEEE 802.1Qci \cite{8021qci} defines frame metering, filtering, and time-based policing mechanisms on a per-stream basis using the stream identification function defined in 802.1CB.

We detail the Time-Aware Shaper (TAS) mechanism defined in IEEE 802.1Qbv~\cite{8021qbv} via the simplified representation of a TSN switch in Figure~\ref{fig:TSNswitch}. The figure presents a scenario in which communication received on one of two available ingress ports (A and B) will be routed to an egress port C. The switching fabric will determine, based on internal routing tables and stream properties, to which egress port a frame belonging to the respective stream will be routed (in our logical representation, there is only one egress port). Each port will have a priority filter that determines which of the available $8$ traffic classes (priorities) of that port the frame will be enqueued in. This selection will be made based on either the priority code point (PCP) contained in the VLAN-tag of 802.1Q frames or the \emph{stream gate instance table} of 802.1Qci, which can be used to circumvent traffic class assignment of the PCP code. As opposed to regular 802.1Q bridges, where the transmission selection sends enqueued frames according to their respective priority, in 802.1Qbv bridges, there is a Time-Aware Shaper (TAS), also called timed-gate, associated with each traffic class queue and positioned before the transmission selection algorithm. A timed-gate can be either in an \emph{open (o)} or \emph{closed (C)} state. When the gate is open, traffic from the respected queue is allowed to be transmitted, while a closed gate will not allow the respective queue to be selected for transmission, even if the queue is not empty. The state of the queues is encoded in a local schedule called Gate-Control List (GCL). Each entry defines a time value and a state (o or C) for each of the $8$ queues. Hence whenever the local clock reaches the specified time, the timed-gates will be changed to the respective open or closed state. If multiple non-empty queues are open at the same time, the transmission selection selects the queue with the highest priority for transmission. 

The Time-Aware Shaper functionality of 802.1Qbv, together with the synchronization protocol defined in 802.1ASrev, enables a global communication schedule that orchestrates the transmission of frames across the network such that real-time constraints (usually end-to-end latencies) are fulfilled. 
The global schedule synthesis has been studied in~\cite{craciunas_rtns_16, SernaRTAS18, Pop16, Frank16} focusing on enforcing deterministic transmission, temporal isolation, and compositional system design for critical streams with end-to-end latency requirements. 

Craciunas et al.~\cite{craciunas_rtns_16} define correctness conditions for generating GCL schedules, resulting in a strictly deterministic transmission of frames with $0$ jitter. Apart from the technological constraints, e.g., only one frame transmitted on a link at a time, the deterministic behavior over TSN is enforced in~\cite{craciunas_rtns_16} through isolation constraints. Since the TAS determines the temporal behavior of entire traffic classes (as opposed to individual frames like in TTEthernet~\cite{steiner10}), the queue state always has to be deterministic. Hence, in~\cite{craciunas_rtns_16}, critical streams are isolated from each other either in the time or space domain by either allowing only one stream to be present in a queue at a time or by isolating streams that are received at the same time in different queues. This condition is called \emph{frame/stream isolation} in~\cite{craciunas_rtns_16}. In~\cite{SernaRTAS18}, critical streams are allowed to overlap to some degree (determined by a given jitter requirement) in the same queue in the time domain, thus relaxing the strict isolation. 

Both approaches enforce that gate states of different scheduled queues are mutually exclusive, i.e., only one gate is open at any time, thus preventing the transmission selection from sending frames based on their assigned traffic class's priority. By circumventing the priority mechanism through the TAS, it is ensured that no additional delay is produced through streams of higher priorities, enforcing thus a highly deterministic temporal behavior.

\section{Timed Efficient Stream Loss-Tolerant Authentication}
\label{sec:TESLA}

TESLA provides a resource efficient way to do asymmetric authentication in a multicast setting~\cite{tesla}. It is described in detail in~\cite{tesla} and~\cite{tesla_rfc}. 

We are considering systems where one end-system wants to send a multicast-signal to multiple receiver end-systems, e.g., periodic sensor data. A message authentication code (MAC), which is appended to each signal, can guarantee authenticity, i.e., that the sender is who he claims to be, and integrity, i.e., that the message has not been altered. The MAC is generated and authenticated by a secret key that all end-systems share (i.e., symmetric authentication). The downside of this approach is that if any of the receiving end-systems is compromised, the attacker would be able to masquerade as the sender by knowing the secret key. In a multicast setting, an asymmetric approach, in which the receivers do not have to trust each other, is preferable.

The traditional asymmetric authentication approach is to use asymmetric cryptography with digital signatures (i.e., private and public keys);  however, as stated in~\cite{tesla_article}, the method is computationally intensive and not well suited for systems with limited resources and strict timing constraints. 

TESLA, however, uses an approach where the source of asymmetry is a time-delayed key disclosure~\cite{tesla}. While this can be implemented with much less overhead, it requires time synchronization between the network nodes. For TSN, the time synchronization is given through the 802.1ASrev protocol.

\begin{figure}[t]
	\centering
\includegraphics[width=0.6\textwidth]{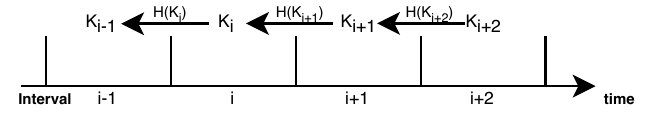}
\caption{TESLA key chain (Adapted from \cite{tesla_article})}
\label{fig:tesla}
\end{figure}

\autoref{fig:tesla} visualizes the TESLA protocol. As described in~\cite{tesla_article}, when using TESLA, time is divided into fixed intervals of length $P_{int}$. At startup a one-way chain of self authenticating keys $K_i$ is generated using a hash function H, where $K_i = H(K_{i+1})$. Each key is assigned to one interval. The protocol is bootstrapped by creating this chain and securely distributing $K_0$ to all receivers~\cite{security_aware_tte}.

Normally in TESLA, as described in~\cite{security_aware_tte}, when a sender sends a message $m$ in the \textit{i-th} interval, it appends to that message: \textit{i}, a keyed-MAC using the key of that interval $K_i$, and a previously used key $K_{i-d}$. Thus, a key remains secret for $d$ intervals. When a receiver receives a message $m$ in the interval $i$ it can not yet authenticate it. It must wait until a message arrives in the interval $i+d$. This message discloses $K_i$, which can be used to decrypt the MAC of $m$ and thus authenticate it. To ensure that $K_i$ itself is valid, we can use any previously validated key. For example, we can check that $H(K_i)=K_{i-1}$, $H(H(K_i))=K_{i-2}$ etc. This makes TESLA also robust to packet loss since any lost keys can be reconstructed from a later key, and any key can always be checked against $K_0$.

Due to the deterministic nature of our schedule, we can make some modifications to the basic TESLA protocol without sacrificing security. The first modification is adopted from
~\cite{security_aware_tte}. Since bandwidth is scarce, we do not release the key $K_{i-d}$ with every message/stream. Instead, it will be released once in its own stream with an appropriate redundancy level. The second modification concerns the TESLA parameter $d$. This parameter is useful in a non-deterministic setting. Since the arrival time of a stream is uncertain, a high value for $d$ makes it more likely that a stream can be authenticated, at the cost of an increased latency. ~\cite{tesla_rfc} However, in our case, we know the exact time a stream will be sent and arrive. Thus, we assume that a stream's keyed-MAC will be generated using the key from the interval it \textit{arrives at the last receiver}. Furthermore, we will release the key $K_{i}$ in the interval, $i+1$ minimizing the latency before a stream can be authenticated.

\section{System Models}
\label{sec:Models}
This section presents the architecture and application models, as well as the threat, security and fault models. Our application model is similar to the one used in related work \cite{security_aware_tte}, but we have extended it to consider TSN networks and the optimization of redundant routing in conjunction with scheduling.

\begin{table}
\centering
\caption{Notations}
\label{tab:notations}
\begin{tabular}{l|l|l}
Description                 & Notation & Unit  \\ 
\hline
Header overhead & $OH$ & Byte\\
Maximum transmission unit & $MTU$ & Byte\\
TESLA key size & $KS$ & Byte\\
TESLA MAC size & $MAC$ & Byte\\
Hyperperiod & H & $\mu s$\\
TSN Network Graph &  $(\mathcal{N},\mathcal{L})$        &       \\
- Nodes                 &$\mathcal{N}=\mathcal{ES}\cup\mathcal{SW}$       &       \\
~ - End-system                  & $e_i \in \mathcal{ES}$ &       \\
~ ~ - Hash computation time & $e_{i}.H$     & $\mu s$    \\
~ - Switch                      & $sw_j \in \mathcal{SW}$ &       \\
- Links                            & $\mathcal{L} \subseteq \mathcal{N} \times \mathcal{N}$         & \\
~ - Network link                 & $l_{a, b}$ &   \\
~ ~ - Link speed                 & $l_{a, b}.s$ & $\mu s$  \\
Application                 & $\lambda_l \in \Lambda$ &   \\ 
- Tuple               & $(\Gamma_l, \mathcal{E}_l)$ &   \\
- Period & $\lambda_l.T$ & $\mu s$\\
- Communication Depth & $\lambda_l.C$ & \\
- Tasks & $t_m \in \mathcal{T}$ & \\
~ - Execution end-system & $t_m.e$ & \\
~ - Worst-case execution time & $t_m.w$ & $\mu s$\\
~ - Period & $t_m.T$ & $\mu s$\\
- Streams & $s_n \in \mathcal{S}$ & \\
~ - Source task & $s_n.t_s$  & \\
~ - Destination tasks & $s_n.T_d$  & \\
~ - Size & $s_n.b$ & Byte\\
~ - Period & $s_n.T$ & $\mu s$\\
~ - Redundancy Level & $s_n.rl$ & \\
~ - Security Level & $s_n.sl$ & \\
~ - MAC generation task & $t^g_{s_n}$ &\\
~ - MAC verification task & $t^{mv}_{s_n}$ &
\\
Security Application                 & $\lambda_l^s \in \Lambda^{sec}$ &   \\ 
- Key release task                 & $t_m^r$ &   \\
- Key verification task                 & $t_m^v$ &   \\
~ - Key source end-system                & $t_m^v.src$ &   \\
- Key stream                 & $s_n^k \in \mathcal{S}_k$ &   \\
\end{tabular}
\end{table}

\subsection{Architecture Model}
We model our TSN network as a directed graph consisting of a set of nodes $\mathcal{N}$ and a set of edges $\mathcal{L}$. The nodes of the graph are either end-systems (ESs) or switches (SWs): \ensuremath{\mathcal{N}=\mathcal{ES}\cup\mathcal{SW}}. The edges $\mathcal{L}$ of the graph represent the network links. 

We assume that all of the nodes in the network are \ac{TSN}-capable, specifically that they support the standards 802.1ASrev \cite{8021asrev} and 802.1Qbv \cite{8021qbv}. Thus we assume the whole network, including the end-systems, to be time-synchronized with a known bounded precision $\delta$. All nodes use the time-aware shaper mechanism from 802.1Qbv to control the traffic flow.

Each end-system $e_i\in\mathcal{ES}$ features a real-time operating system with a periodic table-driven task scheduler. Hash computations, which will be necessary for TESLA operations on that end-system, take $e_i.H$ $\mu s$.

A network link between nodes $n_a \in \mathcal{N}$ and $n_b \in \mathcal{N}$ is defined as $l_{a,b} \in \mathcal{L}$. Since in Ethernet-compliant networks all links are bi-directional and full-duplex, we have that for each $l_{a,b} \in \mathcal{L}$ there is also $l_{b,a} \in \mathcal{L}$. A link $l_{a,b}\in \mathcal{L}$ is defined by a link speed $l_{a,b}.s$.

\begin{figure}[h]
     \centering
     \begin{subfigure}[b]{0.35\textwidth}
         \centering
         \includegraphics[width=\textwidth]{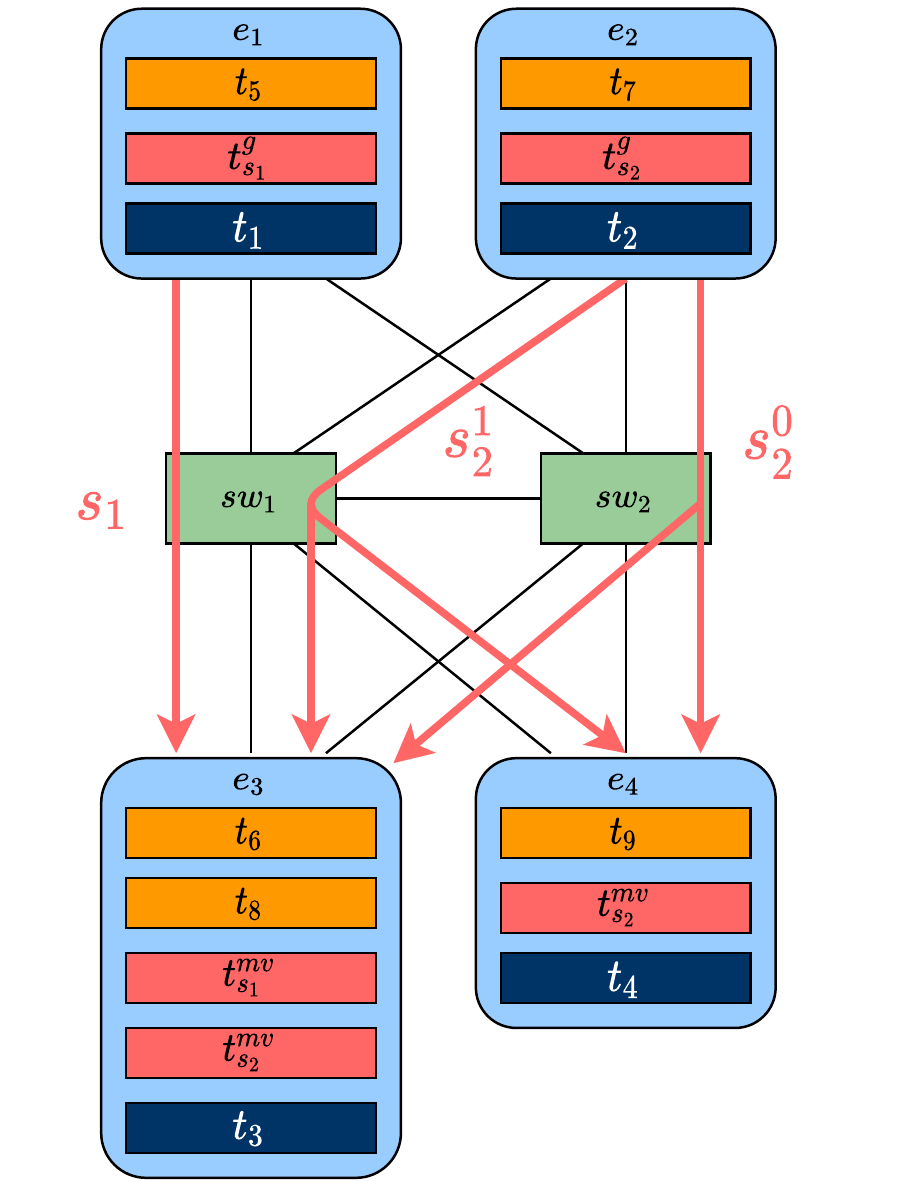}
         \caption{Example Architecture}
         \label{fig:example_arch_and_app:arch}
     \end{subfigure}%
     \vspace{0.5cm}
     \begin{subfigure}[b]{0.45\textwidth}
         \centering
         \includegraphics[width=\textwidth]{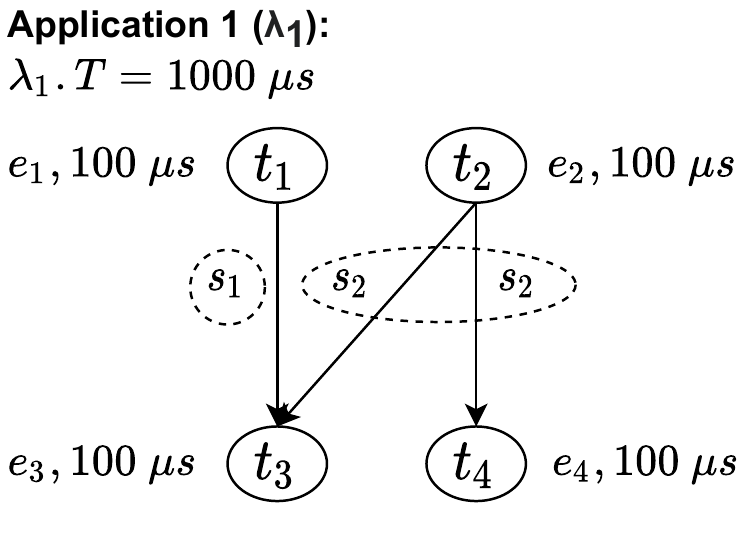}
         \caption{Example Application}
         \label{fig:example_arch_and_app:app}
     \end{subfigure}
     \hfill
    \caption{Example architecture and application models}
    \label{fig:example_arch_and_app}
\end{figure}

\autoref{fig:example_arch_and_app:arch} shows a small example architecture with four end-systems, two switches, and full-duplex links.

\subsection{Application Model}
An application $\lambda_l \in \Lambda$ is modeled as a directed, acyclic graph consisting of a set of nodes representing tasks $\Gamma_l$ and a set of edges $\mathcal{E}_l$ represents a data dependency between tasks. 

A task is executed on a certain end-system $t_m.e$. The worst-case execution time (WCET) of a task is defined by $t_m.w$ $\mu s$. A task needs all its incoming streams (incoming edges in the application graph) to arrive before it can be executed. It produces outgoing streams at the end of its execution time. 
Communication dependencies between tasks that run on the same end-system are usually done via, e.g., shared memory pools or message queues, where the overhead of reading/writing data is negligible and included in the WCET definition of the respective tasks.
Dependencies between tasks on separate end-systems constitute communication requirements and are modeled by streams. A stream in the TSN context is a communication requirement between a sender and one (unicast) or multiple (multicast) receivers. An example application can be seen in \autoref{fig:example_arch_and_app:app}. An application is periodic with a period $\lambda_l.T$, which is inherited by all its tasks and streams.

A stream $s_n$ originates at a source task $s_n.t_s$ and travels to set of destination tasks $s_n.T_d$ (since we consider multicast streams). The stream size $s_n.b$ is assumed to be smaller than the maximum transmission unit (MTU) of the network. Each stream has a redundancy level $s_n.rl$, which determines the amount of required disjunct redundant routes for the stream to take. For each of these routes we model a sub-stream: $s_n^i \in S_{s_n}, 0 \leq i < s_n.rl$
Hereby $S_{s_n}$ is a set containing all sub-streams of $s_n$. This notation is useful to differentiate the different routes a stream takes through the network, and to make sure those routes do not overlap.
A stream also has a binary security level $s_n.sl$ which determines if it is authenticated using TESLA ($sl=1$) or not ($sl=0$). 

We define the hyperperiod $H$ as the least-common multiple of all application periods: $H = lcm(\{\lambda_l.T | \lambda_l \in \Lambda\})$
We define the set $\mathcal{T}$ to contain all tasks and the set $\mathcal{S}$ to contain all streams (including redundant copies).

\subsection{Fault Model}
Reliability models discussed in~\cite{VoicaBahramRedundancy} (e.g., Siemens SN 29500) indicate that the most common type of permanent hardware failures is due to link failures (especially physical connectors) and that ESs and SWs are less likely to fail. These models are complementary to Mean Time to Failure (MTTF) targets established for various safety integrity levels in certification standards such as ISO~26262 for automotive~\cite{VoicaBahramRedundancy}. As mentioned, we assume we know the required redundancy level to protect against permanent link failures.
Our disjoint routing can guarantee the transmission of a stream of RL $n$ despite any $n-1$ link failures. For example, for the routing of $s_2$ with RL 2 in \autoref{fig:example_arch_and_app:arch}, any 1-link failure would still result in a successful transmission.

\subsection{Threat Model}
We use a similar threat model as~\cite{security_aware_tte} and assume that an attacker is capable of gaining access to some end-systems of our system, e.g., through an external gateway or physical access.

We consider that the attackers have the following abilities:
\begin{itemize}
    \item They know about the network schedule and the content of the streams on the network;
    \item They can replay streams sent by other ES;
    \item They can attempt to masquerade as other ES by faking the source address of streams they send;
    \item They have access to all keys released and received by the ES they control;
\end{itemize}

\subsection{Security Model}
We use TESLA to address the threats identified in the previous section, which means that additional security-related models are required. These additional applications, tasks and streams can be automatically generated from a given architecture and application model.

\begin{figure}[htb]
     \centering
     \begin{subfigure}[b]{0.6\textwidth}
         \centering
         \includegraphics[width=\textwidth]{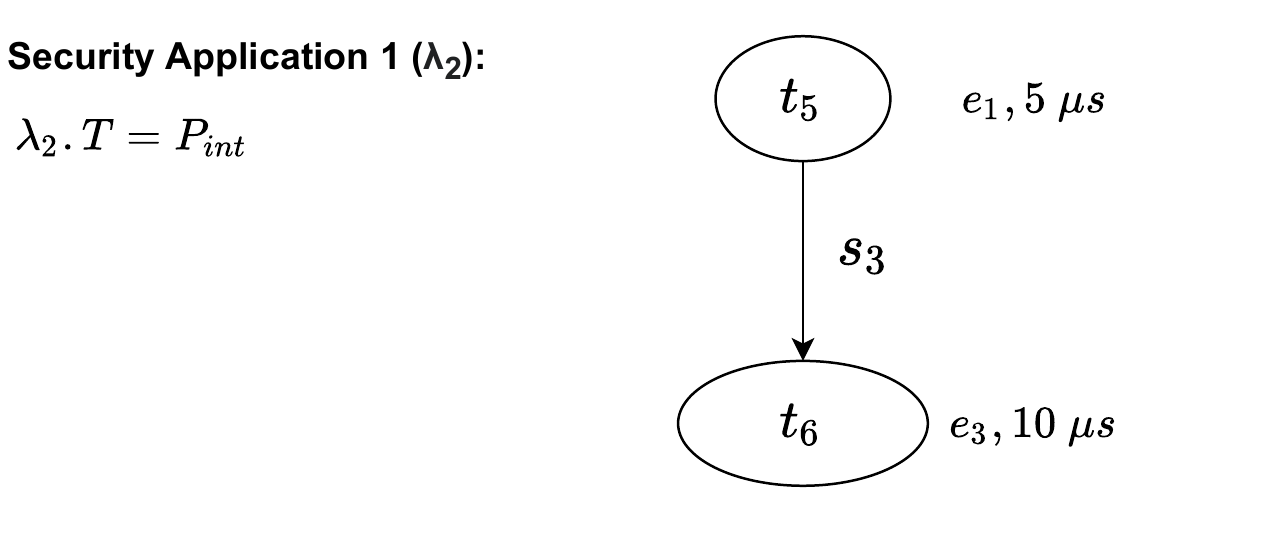}
         \label{fig:example_sec_arch_and_app:arch}
     \end{subfigure}
     \\
     \begin{subfigure}[b]{0.6\textwidth}
         \centering
         \includegraphics[width=\textwidth]{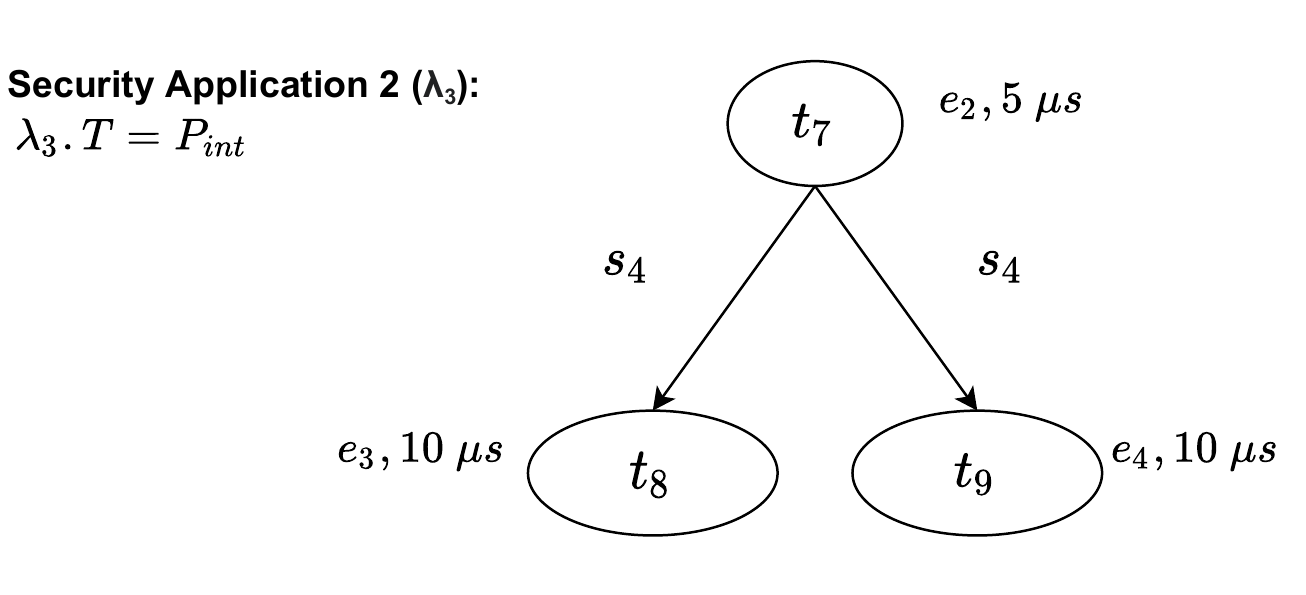}
         \label{fig:example_sec_arch_and_app:app}
     \end{subfigure}
     \hfill
    \caption{Example security model for the applications in \autoref{fig:example_arch_and_app}}
    \label{fig:example_secapps}
\end{figure}

First off, we need to generate, send, and verify a key in each interval for each set of communicating end-systems. We generate a key authentication application $\lambda_s \in \Lambda^{sec}$ for each sender end-system, which is modeled similarly to a normal application as a directed acyclic graph. The period $\lambda_s.T$ is equal to $P_{int}$ (see \autoref{sec:TESLA}) and again inherited by tasks and streams. Each of these application consists of one key release task $t_{m}^r$ scheduled on the sending end-system $e_i$. Additionally, it consists of key verification tasks $t_j^{v}$ on each end-system $e_{j}$ that receives a stream from $e_i$. The release task sends a multicast key-stream $s_{n}^k$ to each of those verification tasks. The redundancy level of a key-stream $s_n^k.rl$ is set to the maximum redundancy level of all streams emitted by $e_i$. The size of a key stream $s_n^k.b$ is equal to the key size $KS$ specified by the TESLA implementation. The security model for our example from \autoref{fig:example_arch_and_app} can be seen in \autoref{fig:example_secapps}.

For a key verification task $t_{m}^{v}.src$ is the end-system $e_i$ whose key this task is verifying. Its execution time is equal to the length of one hash execution on its execution end-system: $t_m^{v}.w = (t_{m}^{v}.e).H$. A key release task's execution time is very short, since the key it releases has already been generated during bootstrapping. We model it to be last half the time of a hash execution: $t_m^{r}.w = \frac{(t_{m}^{r}.e).H}{2}$

Secondly, we need to append MACs to all non-key-streams with $s_n.sl = 1$. Thus, their length increases by the MAC length $MAC$ specified by the TESLA implementation. For each stream $s_n$, a MAC generation task $t^g_{s_n}$ is added to the sender and a MAC validation task $t^{mv}_{s_n}$ to each receiver. Those tasks take the time of one MAC computation on the processing element to execute.

We define the set $\mathcal{T}^n_{kr}$ to contain all key release tasks and $\mathcal{T}^n_{kv}$ to contain all key verification tasks for a given node $n$. Furthermore let $S_{k}$ contain all key streams.

\autoref{fig:example_arch_and_app:arch} shows key release and verification tasks in orange and MAC generation and validation tasks in red.

\autoref{fig:example_secapps} shows the security applications for our example.

\section{Problem Formulation}
\label{sec:Problem_Formulation}
Given a set of applications running on TSN-capable end-systems that are interconnected in a TSN network as described in the architecture, application, and security models in \autoref{sec:Models}, we want to determine a system configuration consisting of:
    \begin{itemize}
        \item an interval duration $P_{int}$ for TESLA operations,
        \item the routing of streams,
        \item the task schedule,
        \item the network schedule as 802.1Qbv \acp{GCL},
    \end{itemize}
such that:
    \begin{itemize}
        \item all deadline requirements of all applications are satisfied.
        \item the redundancy requirements of all streams and the security conditions of TESLA are fulfilled.
        \item the overall latency of applications is minimized.
    \end{itemize}

\subsection{Motivational Example}
\label{subsec:example}
We illustrate the problem using the architecture and application from \autoref{fig:example_arch_and_app}. We have one application \autoref{fig:example_arch_and_app:app} with 4 tasks, 2 streams and a period and deadline of 1000 $\mu s$. The tasks are mapped to the end-systems as indicated in the figure. Stream $s_2$ will be multicast. The size of both streams is 50~B. For TESLA's security requirements, i.e. $s_1.sl = s_2.sl = 1$, we generate two additional security applications (\autoref{fig:example_secapps}).

We have a TSN network with a link speed of 10~Mbit/s and zero propagation delay. Our TESLA implementation uses keys that are 16~B and MACs that are 16~B. A hash computation takes 10~$\mu s$ on every ES.

\begin{figure*}[!htb]
    \vspace{0.25cm}
    \begin{subfigure}[b]{0.98\textwidth}
          \begin{subfigure}[b]{\textwidth}
                  \includegraphics[width=\textwidth]{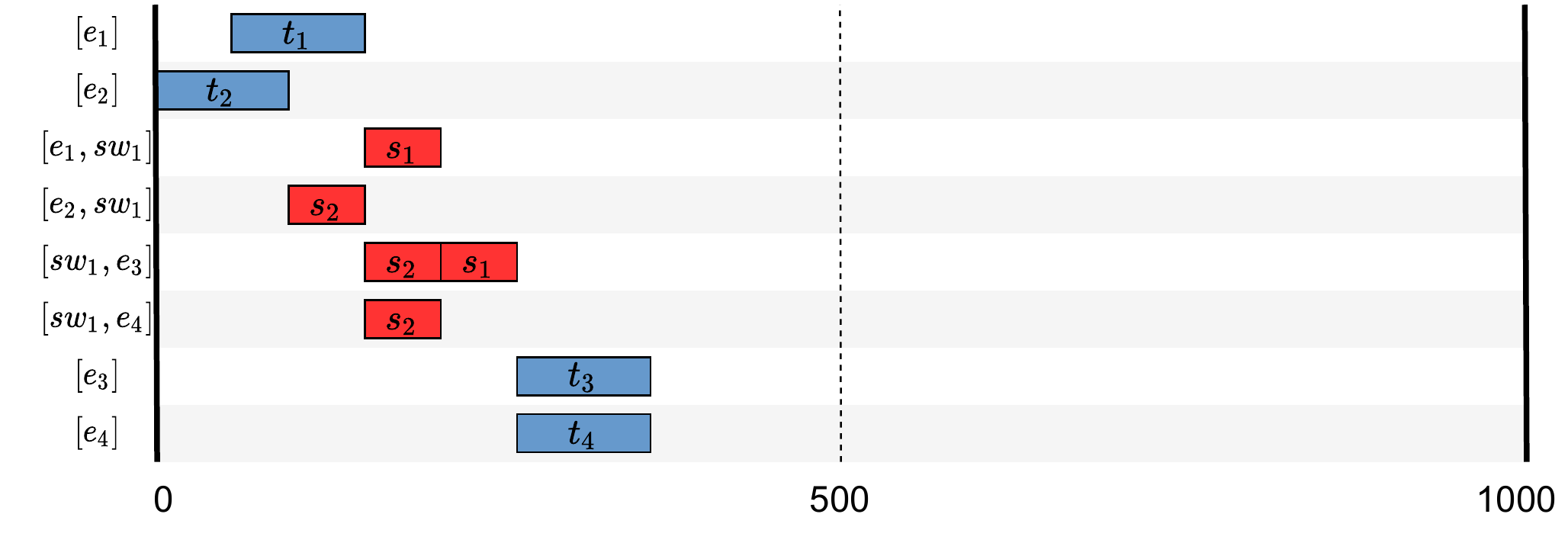}
                  \caption{Schedule without security \& redundancy}\label{fig:example_schedule1}
          \end{subfigure}%
    \end{subfigure}%
    \vspace{0.25cm}
    \begin{subfigure}[b]{0.98\textwidth}
          \begin{subfigure}[b]{\textwidth}
                  \includegraphics[width=\textwidth]{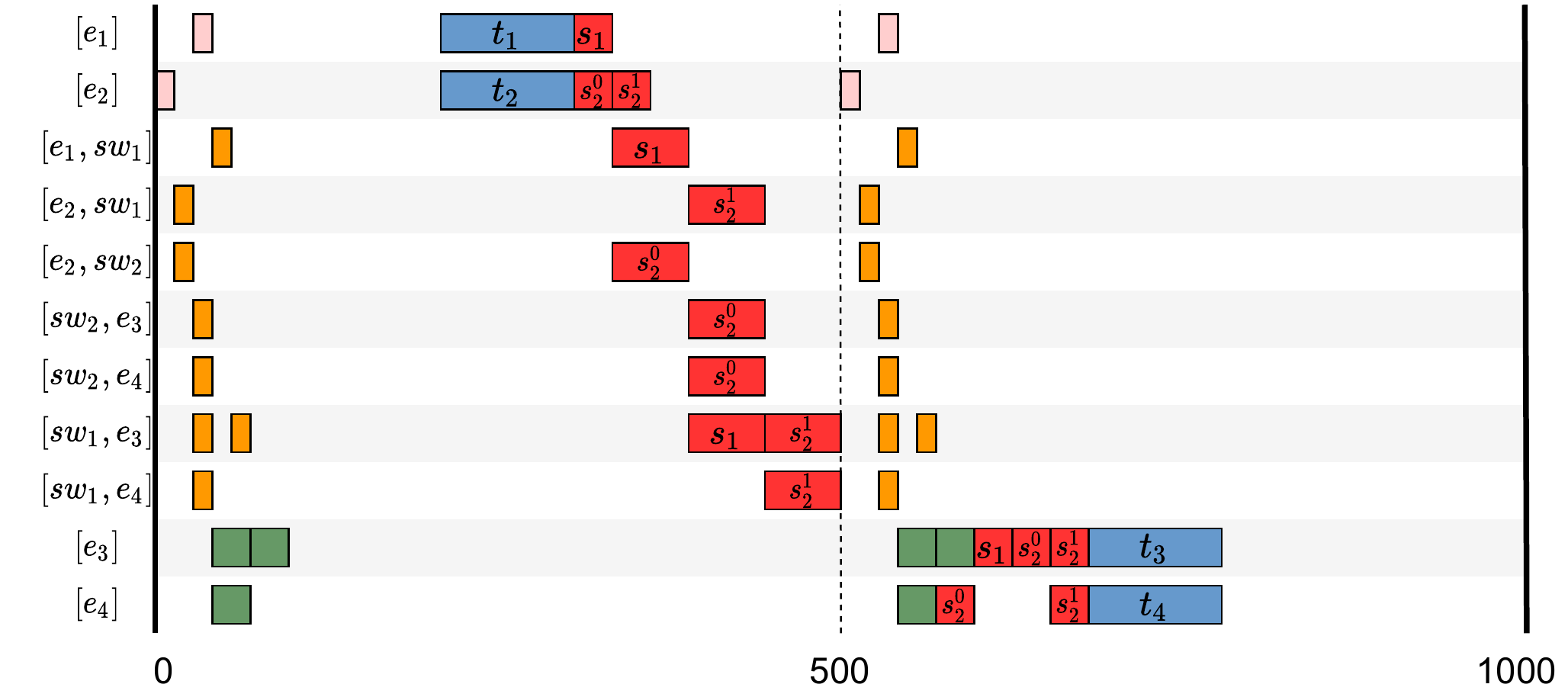}
                  \caption{Schedule with security \& redundancy}\label{fig:example_schedule2}
          \end{subfigure}%
    \end{subfigure}
    \caption{Example solution schedules for the models in \autoref{fig:example_arch_and_app} }\label{fig:example_schedules}
	\vspace*{-10pt}
\end{figure*}

A solution that does not consider the security and redundancy requirements is shown in \autoref{fig:example_schedule1}. With the TSN stream isolation constraint outlined in \autoref{sec:TSN} taken into consideration, the GCLs are equivalent to frame schedules. We depict in \autoref{fig:example_schedule1} the GCLs as a Gantt chart, where the red rectangles show the transmission of streams $s_1$ and $s_2$ on network links, and the blue rectangles show the tasks' execution on the respective end-systems. To guarantee deterministic message transmission in TSN, we have to isolate the frames in the time (or space) domain, leading to the delay of $s_1$ and thus $t_3$. We refer the reader to~\cite{craciunas_rtns_16} for an in-depth discussion on the non-determinism problem and isolation solution in TSN. 

In this paper, we are interested in solutions such as the one in \autoref{fig:example_schedule2}, which considers both the redundancy and security requirements. The black dashed line in the figure separates the TESLA key release intervals, where $P_{int}$ was determined to be 500 $\mu s$. Streams carrying keys are orange, key generation tasks pink, key verification tasks green, and the MAC generation/validation operations on ESs are shown in red. The routing of the non-key streams can be seen in \autoref{fig:example_arch_and_app:arch}. Note how the two redundant copies of $s_2$, $s_2^0$ and $s_2^1$ use non-overlapping paths. 

Of particular importance is the delay incurred by the time-delayed release of keys: tasks $t_3$ and $t_4$ can only be executed after the keys authenticating $s_1$ and $s_2$ have arrived in the second interval, and after key verification and MAC validation tasks have been run. 

Scheduling problems like the one addressed in this paper are NP-hard as they can be reduced to the Bin-Packing problem~\cite{8607243} and may be intractable for large input sizes. In the following sections, we will propose a Constraint Programming (CP) formulation to solve the problem optimally for small test cases, and a heuristic to solve the problem for large test cases.

\section{Constraint Programming Formulation}
\label{sec:Constraint_Formulation}

Constraint Programming (CP) is a technique to solve combinatorial problems defined using sets of constraints on decision variables. For large scheduling problems it becomes intractable to use CP due to the exponential increase in the size of the solution space~\cite{cp_handbook}. 
In order to achieve reasonable runtime performance, we split the problem into $3$ sub-problems which we solve sequentially: (i) finding a route for all streams, (ii) finding $P_{int}$, and (iii) finding the network and task schedule. 

\subsection{Optimizing redundant routing}
The first step of solving the proposed problem is to find a set of (partially) disjoint routes for each stream, depending on the stream's redundancy level. The constraints in this section are inspired by \cite{VoicaBahramRedundancy} and \cite{QuorumcastRouting}. 

We model the stream routes with an integer matrix $X$, where the columns represent streams (including their redundant copies) and rows represent nodes of the network. An entry at the position of a stream $s_n$ and a node $n$ in this matrix referring to a node $m$, represents a link from $m$ to $n$ on the route of stream $s_n$. Alternatively the entry could be $nil$, in which case $n$ is not part of the route.

\begin{table}[]
\centering
\begin{tabular}{|c|c|c|c|}
\hline
X   & \textbf{s1}  & \textbf{s2\_0} & \textbf{s2\_1} \\ \hline
\textbf{ES1} & ES1 & nil   & nil   \\ \hline
\textbf{ES2} & nil & ES2   & ES2   \\ \hline
\textbf{ES3} & SW1 & SW1   & SW2   \\ \hline
\textbf{ES4} & nil & SW1   & SW2   \\ \hline
\textbf{SW1} & ES1 & ES2   & nil   \\ \hline
\textbf{SW2} & nil & nil   & ES2   \\ \hline

\end{tabular}
\caption{Matrix X for example from \autoref{subsec:example}}
\label{tab:routingMatrixX}
\end{table}

Using the matrix $X$, we can construct the route for each stream bottom-up as a tree, by starting at the receiver nodes. See \autoref{tab:routingMatrixX} for the matrix of our example.

To determine the route for each stream $s_n \in \mathcal{S}$, for each node $n \in \mathcal{N}$ we have the following optimization variables:
\begin{itemize}
    \item $x(s_n,n)$ represents an entry of our matrix X. The domain of $x(s_n,n)$ is defined as: $D(x(s_n,n)) = \{m \in \mathcal{N} | l_{m,n} \in \mathcal{L}\} \cup \{n\} \cup \{nil\}$. We refer to $x(s_n,n)$ as the successor of $n$ on the path to the stream sender node.
    \item $y(s_n,n)$ represents the length of the path from $n$ to $s_n.t_s.e$, i.e. the length of the path from node $n$ to the sender node of the stream. $D(y(s_n,n)) = \{i | 0 <= i <= |\mathcal{SW}|+1\}$
\end{itemize}

Furthermore, we define a few helper variables and functions. First off, we define $\mathcal{S}^d$ as the set of all distinct streams, i.e., excluding the redundant copies of streams with redundancy level (RL) greater than one. Additionally we define $\mathcal{S}_{s_d}$ as the set of all redundant copies (including the stream itself) of $s_d$. Then we define the following helper function:
\begin{equation}
    xsum(s_d, n, m) = \sum_{s_d^{\prime} \in \mathcal{S}_{s_d}}^{} (x(s_d^{\prime},n) == m)
\end{equation}
This function allows us, for any given $s_d \in S^d$, to determine the number of redundant copies (including $s_d$ itself) that use the link from $m$ to $n$ (nil is counted as zero).

Then we have the following constraint optimization problem:
 \begin{align*}
     &Minimize: \sum_{s_n \in \mathcal{S}}^{} cost(s_n)\tag{RC1}\label{eq:RC1}\\
 \end{align*}
 where
 \begin{align*}
     &cost(s_n) = \sum_{n \in \mathcal{N} \setminus \{s_n.t_s.e\}}^{} (x(s_n,n) != nil)\tag{RC2}\label{eq:RC2}
 \end{align*}
\textit{s.t.}
\begin{align*}
    &x(s_n,n) \neq nil \Rightarrow y(s_n,n) = y(f, x(s_n,n)) + 1,\tag{R1}\label{eq:R1}\\
    &\phantom{{}=1}\forall s_n \in \mathcal{S},\ n \in \mathcal{N} \setminus \{s_n.t_s.e\}\\%%%
    &x(s_n,m) = nil \Leftrightarrow x(s_n,n) \neq m,\tag{R2}\label{eq:R2}\\
    &\phantom{{}=1}\forall s_n \in \mathcal{S},\ n,m \in \mathcal{N}\\%%%
    &x(s_n,n) \neq nil,\tag{R3.1}\label{eq:R3.1}\\
&\phantom{{}=1}\forall s_n \in \mathcal{S},\ n \in \{t_r.e | t_r \in s_n.T_d\} \\%%%
    &x(s_n,s_n.t_s.e) = s_n.t_s.e,\tag{R3.2}\label{eq:R3.2}\\
    &\phantom{{}=1}\forall s_n \in \mathcal{S}\\%%%
    &x(s_n,n) = nil,\tag{R3.3}\label{eq:R3.3}\\
    &\phantom{{}=1}\forall s_n \in \mathcal{S},\ n \in \mathcal{ES} \setminus \{t_r.e | t_r \in s_n.T_d\} \\%%%
    &y(s_n,s_n.t_s.e) = 0,\tag{R4} \label{eq:R4}\\
    &\phantom{{}=1}\forall s_n \in \mathcal{S}\\%%%
    &\sum_{s_d\in\mathcal{S}^d}\Big ((xsum(s_d,n,m) > 0\Big ) \times \frac{s_d.b}{s_d.T}) \leq [m,n].s, \tag{R5} \label{eq:R5}\\
    &\phantom{{}=1} n,m \in \mathcal{N}\\%%%
    &x(s_n,n) \neq x(s_n^{\prime},n),\tag{R6} \label{eq:R6}\\
    &\phantom{{}=1}\forall s_n \in \mathcal{S},\ s_n^{\prime} \in \mathcal{S}_{s_n} \setminus {s_n},\ n \in \mathcal{N} \setminus \{s_n.t_s.e\},\\%%%
\end{align*}
Please note that \emph{==} and \emph{!=} are boolean expressions that evaluate to 1 if true and to 0 otherwise.

The cost function we are minimizing (\eqref{eq:RC1},\eqref{eq:RC2}) measures the length of the route of each stream.  
\footnote{For some use cases, fully disjoint routes are not necessary. Refer to \autoref{AppendixA} for an updated formulation for this case}

The constraint \eqref{eq:R1} prevents cycles in the route, as defined in~\cite{QuorumcastRouting}. The constraint~\eqref{eq:R2} disallows ``loose ends'', i.e., a node that has a successor/predecessor must have a predecessor/successor itself. Please note that we refer to the successor on the path from receiver to sender, i.e., the predecessor on the route. The constraint~\eqref{eq:R3.1} states that all receivers of a stream have to have a successor. The constraints~\eqref{eq:R3.2},~\eqref{eq:R3.3}, and~\eqref{eq:R4} impose that the sender of the stream has itself as the successor, no other end-system has a successor, and the path length is 0 at the sender node, respectively. The constraint \eqref{eq:R5} restricts the bandwidth usage of each link to be under $100\%$. If multiple copies of the same stream use the same link, only one of them is counted as consuming bandwidth since we assume that streams are intelligently split and merged using IEEE 802.1CB.  The constraint~\eqref{eq:R6} forbids the routes of redundant copies of a stream to overlap at any point.

\subsection{Optimizing $P_{int}$}
\label{subsec:pint}
To set up the TESLA protocol, we need to choose the parameter $P_{int}$.
$P_{int}$ is the duration of one key disclosure interval. It has a big influence on the latency of secure streams and thus on the feasibility/quality of the schedule. 

When choosing $P_{int}$ there is a trade-off between overhead and latency. A small $P_{int}$ reduces the latency of secure streams but necessitates more key generation/verification tasks and key streams. Thus, we want to determine the maximum value of $P_{int}$ for which the latency is still within all deadline bounds. To this end, we formulate constraints inspired by~\cite{security_aware_tte} for which we then determine the optimal solution. This value is used as a constant in the subsequent optimization of the schedule.

We introduce a new notation: For each application $\lambda_l \in \Lambda$ we define $\lambda_l.C$ to be the communication depth, i.e. the length of the longest path in the application graph where only edges with associated secure streams are counted (ES-internal dependencies and non-secure streams are ignored). This gives us a measure of the longest chain of secure communications within the application, which we can use to estimate the amount of necessary TESLA intervals.

Then we have the following formulation:
\begin{align*}
    Maximize: P_{int}\tag{P0}\label{eq_P0}
\end{align*}
s.t.
\begin{align*}
&\forall {\lambda_l \in \Lambda},\quad P_{int}\cdot (\lambda_l.C + 1)\le \lambda_l.T\tag{P1}\label{eq:P1}\\
&H\ mod\ P_{int} = 0 \tag{P2}\label{eq:P2}\\
&P_{int}\ mod\ gcd(\{\lambda_l.T | \lambda_l \in \Lambda\})=0\quad or\\ 
&P_{int}*n = gcd(\{\lambda_l.T | \lambda_l \in \Lambda\}),\quad n\in \mathbb{N}
\tag{P3}\label{eq:P3}
\end{align*}
\vspace{8pt}

The constraint \eqref{eq:P1} guarantees that $P_{int}$ is small enough to accommodate the authentication of all secure streams for all applications. The communication depth $\lambda_l.C$ of an application gives a lower bound of how many TESLA intervals are necessary to accommodate all these streams within the period of the application, since there have to be $n+1$ intervals to accommodate the authentication of $n$ secure streams.

The purpose of the constraints \eqref{eq:P2} and \eqref{eq:P3} is to align the TESLA intervals with the schedule. The ~\eqref{eq:P2} makes $P_{int}$ a divisor of the hyperperiod, while constraint~\eqref{eq:P3} makes $P_{int}$ either a multiple or a divisor of the greatest common divisor of all application periods.

\subsection{Optimizing scheduling}
In this step, we want to find a schedule for all tasks and streams which minimizes the overall latency of streams while fulfilling all constraints imposed by deadlines, TESLA, and TSN. The routes for each stream and $P_{int}$ are given by the previous scheduling steps and assumed constant here.

We define the following integer optimization variables:
\begin{itemize}
    \item $o^s_l$ as the offset of stream $s$ on link or node $l$
    \item $c^s_l$ as the transmission duration of stream $s$ on link or node $l$
    \item $a^s_l$ as the end-time of stream $s$ on link or node $l$
    \item $\varphi^s$ as the index of the earliest interval where stream $s$ can be authenticated on any receiver
    \item $o^t_n$ as the offset of task $t$ (on node $t.e$)
    \item $a^t_n$ as the end-time of task $t$ (on node $t.e$)
\end{itemize}

As an example, let us assume a hyperperiod of 1000us and a stream $s$ with a period of 500us.
$o^s_l=100,\ c^s_l=50,\ a^s_l=150$ would imply that the stream $s$ is scheduled on link $l$ in the following time intervals: (100, 150) and (600, 650).

Furthermore we define several helper variables.
Let $\mathcal{E}^s$ be the set containing all receiver end-systems of stream $s$:
\begin{equation}
    \mathcal{E}^s=\{t.e \mid t \in s.T_d\}\nonumber
\end{equation}
Let $\mathcal{R}^s$ be the set containing all links on the route of stream $s$ as well as sender and receiver nodes: 
\begin{equation}
    \mathcal{R}^s = \{s.t_s.e\} \cup \mathcal{E}^s \cup \{l_{a,b} \mid x(s, b) = a,\ l_{a,b} \in \mathcal{L}\}
\end{equation}

Using these helper functions we define the following constraint-optimization problem for the task and network scheduling step:
\begin{align*}
 &Minimize: \sum_{\lambda_l \in \Lambda}^{} cost(\lambda_l)\tag{CS1}\label{eq:CS1}\\
\end{align*}
where
\begin{align*}
& cost(\lambda_l) = max(\{a^t \mid t \in \Gamma_l\}) - min(\{o^t \mid t \in \Gamma_l\})\tag{CS2}\label{eq:CS2}
\end{align*}
\textit{s.t.}
\begin{align*} 
& cost(\lambda_l) \le \lambda_l.T\tag{S1}\label{eq:S1}\\&\phantom{{}=1} \forall \lambda_l \in \Lambda\\
&o^s_l=c^s_l=a^s_l=0, \tag{S2.1}\label{eq:S2.1}\\
&\phantom{{}=1} \forall s \in \mathcal{S},\ l_{a,b} \in \mathcal{L},\ l_{a,b} \not\in \mathcal{R}^s\\
&o^s_n=c^s_n=a^s_n=0, \tag{S2.2}\label{eq:S2.2}\\
&\phantom{{}=1} \forall s \in \mathcal{S},\ n \in \mathcal{N},\ n \not\in \mathcal{R}^s\\
&o^s_l + c^s_l = a^s_l, \tag{S3.1}\label{eq:S3.1}\\
&\phantom{{}=1}\forall s \in \mathcal{S},\ l_{a,b} \in \mathcal{L},\ l_{a,b} \in \mathcal{R}^s\\
&o^s_n + c^s_n = a^s_n, \tag{S3.2}\label{eq:S3.2}\\
&\phantom{{}=1} \forall s \in \mathcal{S},\ n \in \mathcal{N},\ n \in \mathcal{R}^s\\
&c^s_l = \left \lceil \dfrac{s.b}{l.s} \right \rceil, \tag{S4.1}\label{eq:S4.1}\\
&\phantom{{}=1}\forall s \in \mathcal{S},\ l_{a,b} \in \mathcal{L},\ l_{a,b} \in \mathcal{R}^s\\
&c^s_n = n.H, \tag{S4.2}\label{eq:S4.2}\\
&\phantom{{}=1}\forall s \in \mathcal{S},\ n \in \mathcal{N} \cap \mathcal{R}^s, s.secure==1\\
\end{align*}

The constraint \eqref{eq:S1} sets the deadline for the completion of an application to its period. The constraints \eqref{eq:S2.1} and \eqref{eq:S2.2} set all optimization variables to zero for every stream, for all nodes and links not part of its route. For all other links and nodes constraints, \eqref{eq:S3.1} and \eqref{eq:S3.2} set the end-time to be the sum of offset a length. For each link on the route of a stream constraint \eqref{eq:S4.1} sets the length to be the byte-size of the stream divided by the link-speed. In constraint \eqref{eq:S4.2} the length of secure streams on end-systems is set to the length of one hash-computation on that end-system,  approximating the duration of MAC generation/verification.

\begin{align*}
&\varphi^s > \left \lfloor \frac{a^s_n}{P_{int}} \right \rfloor, \tag{S5}\label{eq:S5}\\
&\phantom{{}=1}\forall s \in \mathcal{S},\ l_{a,b} \in \mathcal{L}\cap\mathcal{R}^s,\ b \in \mathcal{E}^s,\ s.secure==1\\
&o^s_n \ge a^{t_{key}} + \varphi^s * P_{int}\tag{S6}\label{eq:S6}\\&\phantom{{}=1}\forall s \in \mathcal{S},\ n \in \mathcal{E}^s, s.secure=1\\
&\phantom{{}=1}\forall t_{key} \in T_{kv}^n\\
&a^s_{l_{a,b}} \le o^s_{l_{b,c}}\tag{S7.1}\label{eq:S7.1}\\&\phantom{{}=1}\forall s \in \mathcal{S},\ l_{b,c} \in \mathcal{L} \cap \mathcal{R}^s\\
&\phantom{{}=1}a=x(s,b)\\%
&a^s_{a} \le o^s_{l_{a,b}}\tag{S7.2}\label{eq:S7.2}\\&\phantom{{}=1}\forall s \in \mathcal{S},\ s.secure==1,\\
&\phantom{{}=1}l_{a,b} \in \{l_{a,b} \mid l_{a,b} \in \mathcal{L} \cap \mathcal{R}^s,\ a=s.t_s.e\}\\%
&a^s_{l_{a,b}} \le o^s_{b}\tag{S7.3}\label{eq:S7.3}\\&\phantom{{}=1}\forall s \in \mathcal{S},\ s.secure==1,\\
&\phantom{{}=1}l_{a,b} \in \{l_{a,b} \mid l_{a,b} \in \mathcal{L} \cap \mathcal{R}^s,\ b \in \mathcal{E}^s\}\\%
\end{align*}

In constraint \eqref{eq:S5} the earliest authentication interval for a stream $\varphi^s$ is bound to be after the latest interval where the stream is transmitted. In constraint \eqref{eq:S6} the start time of the stream on any receiver end-system is then bound to be greater or equal to the start time of that interval plus the end-time of the necessary preceding key verification task.
The constraints \eqref{eq:S7.1}, \eqref{eq:S7.2} and \eqref{eq:S7.3} make sure that every stream is scheduled consecutively along its route. Hereby constraint \eqref{eq:S7.1} enforces the precedence among two links, \eqref{eq:S7.2} among the MAC generation on the sender and the first link and \eqref{eq:S7.3} among the last link and the following MAC verification.

\begin{align*}
&(\alpha\times s_1.T+a^{s_1}_l \le \beta \times s_2.T + o^{s_2}_l) \lor(\beta \times s_2.T + a^{s_2}_l <= \alpha \times s_1.T + o^{s_1}_l)\tag{S8}\label{eq:S8}\\
&\phantom{{}=1}\forall s_1, s_2 \in \mathcal{S}, s_1 \neq s_2, \forall l \in \mathcal{R}^s_1 \cap \mathcal{R}^s_2,\\
&\phantom{{}=1}\forall \alpha \in \{0, ..., lcm(s_1.T, s_2.T) / s_1.T\},\phantom{{}=1}\forall \beta \in \{0, ..., lcm(s_1.T, s_2.T) / s_2.T\}\\%
&(\alpha \times s_2.T+o^{s_2}_{l_{b,c}} <= \beta \times s_1.T + o^{s_1}_{l_{a_1,b}}) \lor(\beta \times s_1.T + o^{s_1}_{l_{b,c}} <= \alpha \times s_2.T + o^{s_2}_{l_{a_2,b}})\tag{S9}\label{eq:S9}\\
&\phantom{{}=1}\forall s_1, s_2 \in \mathcal{S}, s_1 \neq s_2, \forall l \in \mathcal{R}^s_1 \cap \mathcal{R}^s_2,\\
&\phantom{{}=1}a_1 = x(s_1, b),\ a_2 = x(s_2,b),\\
&\phantom{{}=1}\forall \alpha \in \{0, ..., lcm(s_1.T, s_2.T) / s_1.T\},\phantom{{}=1}\forall \beta \in \{0, ..., lcm(s_1.T, s_2.T) / s_2.T\}
\end{align*}

The constraint \eqref{eq:S8} prevents any streams from overlapping on any nodes or links. Furthermore, constraint \eqref{eq:S9} guarantees that for each link connected to an output port of a switch, the frames arriving on all input ports of that switch that want to use this output port cannot overlap in the time domain. This is the frame isolation necessary for determinism in our TSN configuration, which is further explained in \cite{craciunas_rtns_16}.

\begin{align*}
    &o^t + t.w = a^t\tag{T1}\label{eq:T1}\\&\phantom{{}=1}\forall t \in \mathcal{T}\\%
    &a^t \le o^s_{t.e} \tag{T2.1}\label{eq:T2.1}\\&\phantom{{}=1}\forall t \in \mathcal{T},\ s \in \mathcal{S},\ s.t_s = t,\ s.secure==1\\%
    &a^t \le o^s_{l_{a,b}} \tag{T2.2}\label{eq:T2.2}\\&\phantom{{}=1}\forall t \in \mathcal{T},\ s \in \mathcal{S},\ s.t_s = t,\ s.secure==0\\&\phantom{{}=1}\forall l_{a,b} \in \mathcal{L}\cap \mathcal{R}^s,\ a==t.e\\%
    &a^s_{t.e} \le o^t \tag{T3.1}\label{eq:T3.1}\\&\phantom{{}=1}\forall t \in \mathcal{T},\ s \in \mathcal{S},\ t \in s.T_d,\ s.secure==1\\%%
    &a^s_{l_{a,b}} \le o^t \tag{T3.2}\label{eq:T3.2}\\&\phantom{{}=1}\forall t \in \mathcal{T},\ s \in \mathcal{S},\ t \in s.T_s,\ s.secure==0\\&\phantom{{}=1}\forall l_{a,b} \in \mathcal{L}\cap \mathcal{R}^s,\ b \in \mathcal{E}^s\\
    &(\alpha\times t_1.T+a^{t_1} \le \beta \times t_2.T + o^{t_2}) \quad\lor\tag{T4}\label{eq:T4}\\&(\beta \times t_2.T + a^{t_2} \le \alpha \times t_1.T + o^{t_1})\\
    &\phantom{{}=1}\forall t_1, t_2 \in \mathcal{T},\ t_1 \neq t_2,\\
    &\phantom{{}=1}\forall \alpha \in \{0, ..., lcm(t_1.T, t_2.T) / t_1.T\},\\&\phantom{{}=1}\forall \beta \in \{0, ..., lcm(t_1.T, t_2.T) / t_2.T\}\\%
    &(\alpha\times t.T+a^{t} \le \beta \times s.T + o^{s}_{t.e}) \quad\lor\tag{T5}\label{eq:T5}\\&(\beta \times s.T + a^{s}_{t.e} \le \alpha \times t.T + o^{t})\\
    &\phantom{{}=1}\forall t \in \mathcal{T},\ s \in \mathcal{S}, s.secure==1, t.e \in \mathcal{R}^s\\
    &\phantom{{}=1}\forall \alpha \in \{0, ..., lcm(t.T, s.T) / t.T\},\\&\phantom{{}=1}\forall \beta \in \{0, ..., lcm(t.T, s.T) / s.T\}\\%
\end{align*}

The constraint \eqref{eq:T1} sets the end-time of a task to be the sum of offset and length.
The constraints \eqref{eq:T2.1} and \eqref{eq:T2.2} model the dependency between a task and all its outgoing streams: such streams may only start after the task has finished.
Similarly, constraints \eqref{eq:T3.1} and \eqref{eq:T3.2} model the dependency between a task and its incoming streams: such a task may only start after all incoming streams have arrived.
Finally, constraint \eqref{eq:T4} prevents any two tasks from overlapping, while constraint \eqref{eq:T5} prevents a task from overlapping with a MAC generation/verification operation.

\section{Metaheuristic Formulation}
\label{sec:metaheuristic}

As mentioned in \autoref{sec:Constraint_Formulation}, the scheduling problem addressed in this paper is NP-hard. As a consequence, a pure CP formulation solved using a CP solver is not tractable for large problem sizes. Hence, in this section, we propose a metaheuristic-based strategy, which aims to find good solutions (without the guarantee of optimality) in a reasonable time, even for large test cases.

An overview of our strategy is presented in \autoref{alg:outer}. We use a Simulated Annealing (SA) metaheuristic \cite{SimulatedAnnealing} to find solutions $\Phi = (\mathcal{R}, \Sigma)$, consisting of a set of routes $\mathcal{R}$ and a schedule $\Sigma$. As an input, we provide our architecture model $(\mathcal{N}, \mathcal{L})$ and the application model $\Lambda$. SA randomly explorers the solution space in each iteration by generating ```neighbors'' of the current solution using design transformations (or ``moves''). We consider both routing and scheduling-related moves, and the choice is controlled by a $p_{rmv}$ parameter that gives the probability of a routing move. To measure the quality of a solution  we use a cost function with two parameters $a$ and $b$ which are factors for punishing overlap of redundant streams and missed deadlines for applications, respectively. While we always accept better solutions, the central idea of Simulated Annealing is to also accept worse solutions with a certain probability in order to not get stuck in local optima~\cite{burke2005search}.

\autoref{alg:outer} shows the main loop of the heuristic. We start out with an initial solution, a cost value, and a positive temperature. (line 2-4). Then, we repeat the steps described below until a stopping criterion like a time- or iteration-limit is met. We create a slight permutation of the current solution $\Phi$ by using the $RandomNeighbour$ function (line 6). We calculate the cost of the new solution (line 7) and a delta of the new and old cost (line 8). Now, if the delta is smaller than 0, i.e., if $\Phi_{new}$ is a better solution than $\Phi$, we choose $\Phi_{new}$ as the current solution (line 10-12). Alternatively, the new solution is also accepted if a random chosen value between 0 and 1 is smaller than the value of the acceptance probability function $e^{-\frac{\delta}{t}}$. This acceptance probability will decrease with the temperature over time and is also influenced by $\delta$, which gives a measure of how much worse the new solution is. Finally, since we will occasionally accept worse solutions, we keep track of the best cost achieved overall and adjust it if necessary (line 12-14).

\begin{algorithm}[t]
\SetAlgoLined
\SetKwInOut{Input}{input}
\SetKwInOut{Output}{output}
\SetKwProg{Fn}{Function}{}{}
\Fn{$heuristic(\mathcal{N}, \mathcal{L}, \Lambda, T_{start}, \alpha, k, p_{rmv}, a, b, w)$}{
$\Phi_{best}$ = $\Phi$ = InitialSolution($\mathcal{N}, \mathcal{L}, \Lambda$, $k$)\; \label{alg:line:outerloop_initial_solution}
$c_{best}$ = c = Cost($\Phi$, $a$, $b$)\;
t = $T_{start}$\;
\While{stopping-criterion not True}
{
$\Phi_{new}$ = RandomNeighbour($\Phi, p_{rmv}$)\;
$c_{new}$ = Cost($\Phi_{new}$, $a$, $b$)\;
$\delta$ = $c_{new} - c$\;
\uIf{$\delta < 0$ \textbf{or} random[0,1) $< e^{-\frac{\delta}{t}}$}
{
$\Phi$ = $\Phi_{new}$\;
$c$ = $c_{new}$\;
\uIf{$c_{new} < c_{best}$}
{
    $\Phi_{best}$ = $\Phi_{new}$\;
    $c_{best}$ =  $c_{new}$
}
}
t = $t * \alpha$\;
}
\KwRet $\Phi_{best}$\;
}
\caption{Simulated Annealing Metaheuristic}
\label{alg:outer}
\end{algorithm}

\subsection{Precedence graph}

\begin{figure}[b]
    \centering
    \includegraphics[width=0.6\textwidth]{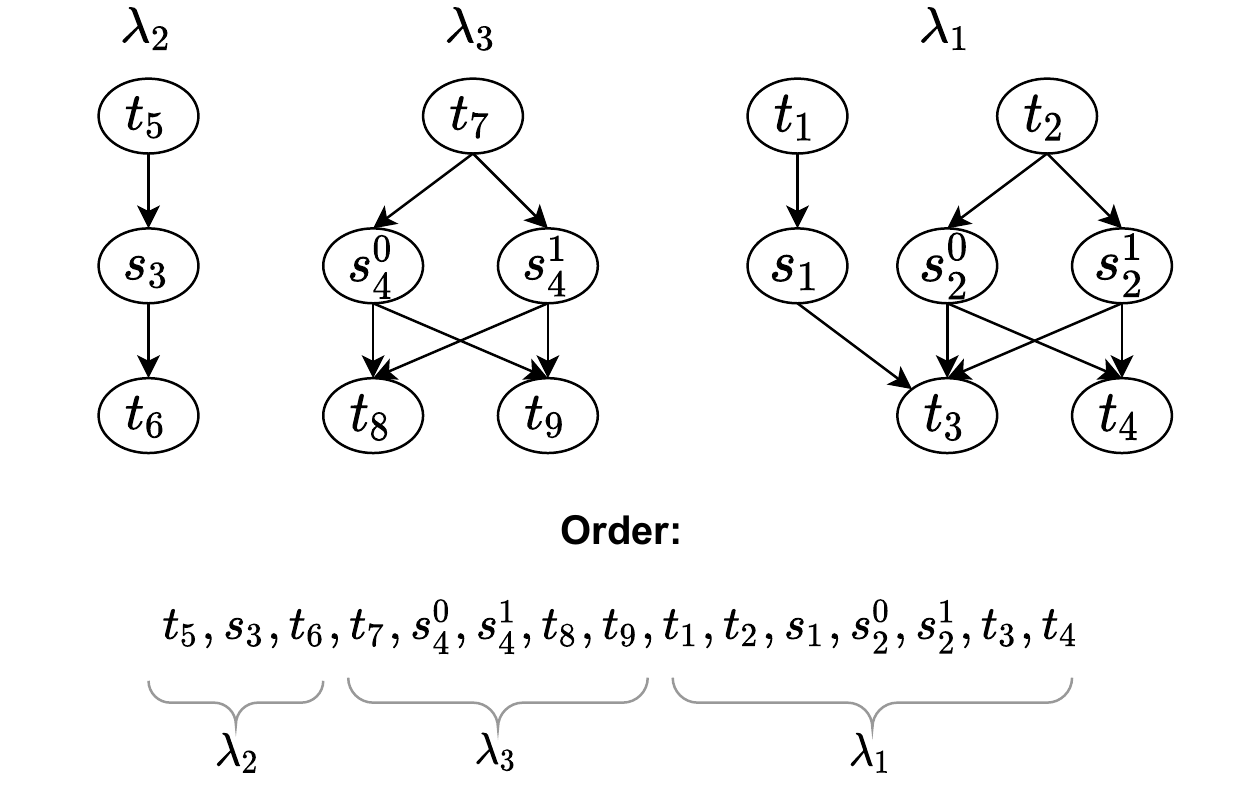}
    \caption{Example precedence graph with associated order}
    \label{fig:task_graph}
\end{figure}

We introduce a helper data structure in the form of a precedence graph. A precedence graph is a collection of special DAGs, one for each application. These DAGs are expanded versions of the DAGs from the application model. Here, streams are modeled as nodes instead of edges, and each redundant copy of a stream has its own node. See \autoref{fig:task_graph} for an example. This data structure helps to model all the dependencies between tasks and streams in the scheduling algorithm. Additionally, we will use the set of all topological orders of this graph as our solution space for the scheduling step. An order can be seen as a scheduling priority assignment that respects all precedence constraints.

\subsection{Initial solution}

\begin{algorithm}
\SetAlgoLined
\SetKwInOut{Input}{input}
\SetKwInOut{Output}{output}
\SetKwProg{Fn}{Function}{}{}
\Fn{InitialSolution($\mathcal{N}, \mathcal{L}, \Lambda, k, w$)}
{
\tcp{routing}
\ForEach{$s \in \mathcal{S}$}
{
\ForEach{$e_{r} \in \{t_r.e | t_r \in s.T_d\}$}
{
\uIf{IsFirstCopyOfStream($s$)}
{
$K_s^{e_r}$ = ShortestPaths($s.t_s.e, e_r, k, \mathcal{N}, \mathcal{L}$)\;
}
\uElse{
$K_s^{e_r}$ = ShortestPathsWeighted($s.t_s.e, e_r, k, \mathcal{N}, \mathcal{L}, w$)\;
}
$\Phi.R^s$ = ShortestPath($K_s^{e_r}$) \;
}
}
\tcp{schedule}
P = CreatePrecedenceGraph($\Lambda$)\;
\ForEach{$\lambda_s \in \Lambda^{sec}$}
{
    $O$ = $O\ \cup$ TopologicalOrder($\lambda_s, P$)\;
}
\ForEach{$\lambda_n \in \Lambda \setminus \Lambda^{sec}$}
{
    $O$ = $O\ \cup$ TopologicalOrder($\lambda_n, P$)\;
}
$\Phi.K = K; \Phi.P = P; \Phi.O = O$\;
$\Phi.\Sigma$ = Schedule($O, \Phi.\mathcal{R}$)\;
\KwRet $\Phi$\;
}

\caption{InitialSolution}
\label{alg:init_sol}
\end{algorithm}

In the beginning, we create an initial solution $\Phi$ from the given architecture and application model. A solution is a tuple $(\mathcal{R}, \Sigma)$ consisting of a set of routes $\mathcal{R}$ and a schedule $\Sigma$. \autoref{alg:init_sol} details the function to find the initial solution.

To find an initial set of routes, we iterate through all streams and all pairs of sender and receiver ES (lines 2-3). For each such pair, we calculate and store $k$ shortest paths for the given topology (line 5). For each redundant copy of a stream beyond the first, we calculate the shortest path in a weighted graph, where we weight all link used by previous copies with $w$ instead of 1 (line 7). For the initial solution, we choose the shortest path for each pair (line 8). Note that our k-shortest-path algorithm only generates paths without repeated nodes that do not traverse any end-system.

To find an initial schedule, we have to create the precedence graph $P$ (line 11) and decide an order $O$ of this graph.

For the initial solution, we construct an order on the level of applications, i.e., we avoid interleaving nodes of different applications. We prioritize key applications (lines 12-14) before other (normal) applications (lines 15-17). This order is consequently used to create a schedule (line 19). See \autoref{fig:task_graph} for an example order.

\subsection{Neighbourhood function}
\label{sec:neighbour}

\begin{algorithm}
\SetAlgoLined
\SetKwInOut{Input}{input}
\SetKwInOut{Output}{output}
\SetKwProg{Fn}{Function}{}{}
\Fn{RandomNeighbour($\Phi, p_{rmv}$)}
{
p = $random[0,1]$\;
\eIf{$p < p_{rmv}$} 
{
   s = RandomStream($\Phi$)\; \label{alg:line:rand_stream}
   $e_{r}$ = RandomReceiver(s)\; \label{alg:line:rand_recv}
   $\Phi.\mathcal{R}^s$ = RandomPath($\Phi.K^{e_r}_s$)\; \label{alg:line:new_path}
   
}
{
    $d_1$ = RandomNormalApplication($\Phi$)\; \label{alg:line:rand_app}
    $d_2$ = RandomNormalApplication($\Phi$)\; \label{alg:line:rand_app2}
    $\Phi.O$ = SwitchSchedulingOrder($d_1$, $d_2$, $\Phi.O$)\; \label{alg:line:switch_order}
    $\Phi.\Sigma$ = Schedule($\Phi.O, \Phi.\mathcal{R}$)\; \label{alg:line:neighb_schedule}
    $\Phi.\Sigma$ = OptimizeLatency($\Phi.\Sigma$, $\Phi.P$)\; \label{alg:line:optimize}
}
\KwRet $\Phi$\;
}

\caption{RandomNeighbour}
\label{alg:neighbour}
\end{algorithm}

The neighbourhood function $RandomNeighbour(\Lambda, p_{rmv})$ is detailed in \autoref{alg:neighbour}. It is used during Simulated Annealing to create a slight permutation of a given solution/candidate $\Phi$. It contains two fundamental moves: Changing the routing $\Lambda.\mathcal{R}$ or changing the schedule $\Lambda.\Sigma$. Which move is taken is decided randomly (line 3). The parameter $p_{rmv}$ influences how likely it is that the routing move is taken, e.g., $p_{rmv}=0.5$ would result in a probability of 50\%.

A routing move consists of choosing a random stream $s$ out of the set of all streams (\autoref{alg:line:rand_stream}), choosing a random receiver $e_{r}$ out of all receivers of that stream (\autoref{alg:line:rand_recv}) and then assigning a random path out of the set of k-shortest-paths calculated during the creation of the initial solution (\autoref{alg:line:new_path}).

A scheduling move consists of choosing two random normal (non-key) applications $d_1$ and $d_2$ (lines \ref{alg:line:rand_app},\ref{alg:line:rand_app2}), switching their order $O$ in the precedence graph $P$ (\autoref{alg:line:switch_order}) and recalculating the schedule (\autoref{alg:line:neighb_schedule}). Whenever a new schedule is calculated, we also optimize its latency (\autoref{alg:line:optimize}). This is further explained in \autoref{sec:optimizelatency}.

\subsection{Cost function}
\label{sec:cost}
The cost function is used in the simulated annealing metaheuristic to evaluate the quality of a solution. A lower cost means a better solution. \autoref{alg:cost} shows how our cost function is calculated. It consists of two components: a routing cost $c_{route}$ and a schedule cost $c_{sched}$. The routing cost is the sum of the number of overlaps of redundant stream (one for each stream for each link) which is punished with a factor $a$ and the total accrued length of all routes. The schedule cost is the sum of the number of infeasible applications, which is punished with a factor $b$, and the total sum of all application latencies (distance between start-time of first task and end-time of the last task). The factors $a$ and $b$ should be sufficiently high such that solutions with less overlap and infeasible applications are preferred.

\begin{algorithm}
\SetAlgoLined
\SetKwInOut{Input}{input}
\SetKwInOut{Output}{output}
\SetKwProg{Fn}{Function}{}{}
\Fn{Cost($\Phi$, $a$, $b$)}
{
$c_{route}$ = $a$ * Overlaps($\Phi.\mathcal{R}$) + Length($\Phi.\mathcal{R}$)\;

$c_{sched}$ = $b$ * Infeasible($\Phi.\Sigma$) + Latency($\Phi.\Sigma$)\;

\KwRet $c_{route} + c_{sched}$\;
}

\caption{Cost}
\label{alg:cost}
\end{algorithm}

\subsection{ASAP list scheduling}
To calculate a schedule for a given precedence graph with associated order and routing, we use an ASAP list-scheduling heuristic \cite{SinnenTaskSchedulingForParallelSystems}, which schedules each node of the precedence graph in the given order.

The algorithm, presented in \autoref{alg:schedule}, starts by iterating through each entry $n$ of the given order $O$ (line 2). An entry may either be a task or a stream. For each entry, we determine where it will be scheduled and create an indexable list $L$ with all these locations (\autoref{alg:line:sched:route}). For a task, that set would contain just one end-system, while for a stream, it may contain many links (which are synonymous to an output port of a switch/ES) and also multiple end-systems, if the stream is secure, thus requiring MAC generation/verification.

Using these locations we also create a set of blocks (\autoref{alg:line:sched:blocks}). A block $b$ is a tuple $(e,l,o,\underline{o},\overline{o},prev,next)$ which is associated to an entry $e$ (task/stream) and a location $l$ (node/link). $o$ represents the block offset. $\underline{o}$ and $\overline{o}$ are parameters representing a lower and upper bound on the offset, which are used during the algorithm. The set $B$ is implemented as a linked list, where $prev$ and $next$ are references to neighboring blocks on the route $L$. Note that in the case of multicast streams $next$ could contain references to multiple blocks.

\begin{algorithm}
\SetAlgoLined
\SetNoFillComment
\SetKw{Break}{break}
\SetKwProg{Fn}{Function}{}{}
\SetKwInOut{Input}{input}
\SetKwInOut{Output}{output}
\Fn{Schedule($P, \mathcal{R}$)}{
    \ForEach{$n \in O$} 
    {
    L = GetRoute(n, $\mathcal{R}$)\; \label{alg:line:sched:route}
    B = CreateBlocks(n, L)\; \label{alg:line:sched:blocks}
    l = L[0]\;
    i = 0\;
    \While{true} {
        b = B[l]\;
        $b.\underline{o}$ = CalculateLowerBound(n, b, $P$, $\mathcal{R}$)\;
        $o$ = EarliestOffset(b, l)\;
        \BlankLine
        \uIf{$o$ == $\infty$}{
            \KwRet false\;
        }
        \uElseIf{$o \leq b.\overline{o}$}
        {
            $b.o$ = $o$\;
            \ForEach{$g \in b.next$}
            {
                \uIf{IsBlockOnLink(g)}
                {
                    $g.\overline{o}$ = LatestQueueAvailableTime(g, $o$)\;
                }
            }
            i = i + 1\;
            \eIf{i $<$ len(L)}
            {
                l = L[i]\;
            }
            {
                \Break\;
            }
        }
        \Else
        {
            g = b.prev\;
            $g.\underline{o}$ = EarliestQueueAvailableTime(b, $o$)\;
            l = b.prev.l\;
            i = L.indexOf(l)\;
        }
    }
    
    $\Sigma$ = UpdateSchedule(B)\; \label{alg:line:update_schedule}
    
    }
    
    \KwRet $\Sigma$\;
}
\caption{Scheduling - ASAP Heuristic}
\label{alg:schedule}
\end{algorithm}

We now iterate over all these blocks (lines 7-8). For each block we begin by calculating the lower bound on the offset (line 9)\footnote{The algorithm can be found in \autoref{sec:appendixb} \label{fn:1}}. Usually, this lower bound is going to be the end-time (offset+length) of the block on the previous link, making sure that a stream is scheduled consecutively along its route. The first block is the maximum of all end-times of the last blocks of the predecessors of the current entry $n$ in the precedence graph. For example for application $\lambda_1$ in \autoref{fig:task_graph}, the lower bound of the offset of the block of $t_3$ would be set to the maximum of the end-times of the last blocks of $s_1$, $s_2^0$ and $s_2^1$.

Also, for a secure stream, for all blocks on receiver ESs (i.e., MAC validation tasks), the lower bound is set to the end-time of the corresponding key verification task in the TESLA interval after the stream was received on the ES, since, according to the TESLA security condition, the stream can only be authenticated from that point on.

In the next step, the earliest possible offset for the current block is calculated (line 10). This function returns the earliest offset greater or equal to the lower bound within the feasible region. For more detail see \autoref{sec:earliestoffset}.

If such an offset is found and it is smaller than or equal to the upper bound, we can assign it to the block (line 14). We then iterate through each of the following blocks and set their upper bound to the latest point in time when their node is available and has been since the offset (line 15-18). This is done to fulfill the TSN constraint which forbids different streams to interleave within a queue (c.f.~\cite{raagard}, \cite{craciunas_rtns_16} for a more detailed explanation).

If such an offset is found but it is larger than the upper bound, it is impossible to schedule the block while the port is still available, i.e., without it interleaving with other streams (line 25). Consequently, we have to backtrack and schedule the previous block at a later time. Therefore we set the lower bound \textit{of the previous block} to the earliest time when the current port is available and remains so until the offset (line 26-27).

\begin{figure}
     \centering
     \begin{subfigure}[b]{0.75\textwidth}
         \centering
         \includegraphics[width=\textwidth]{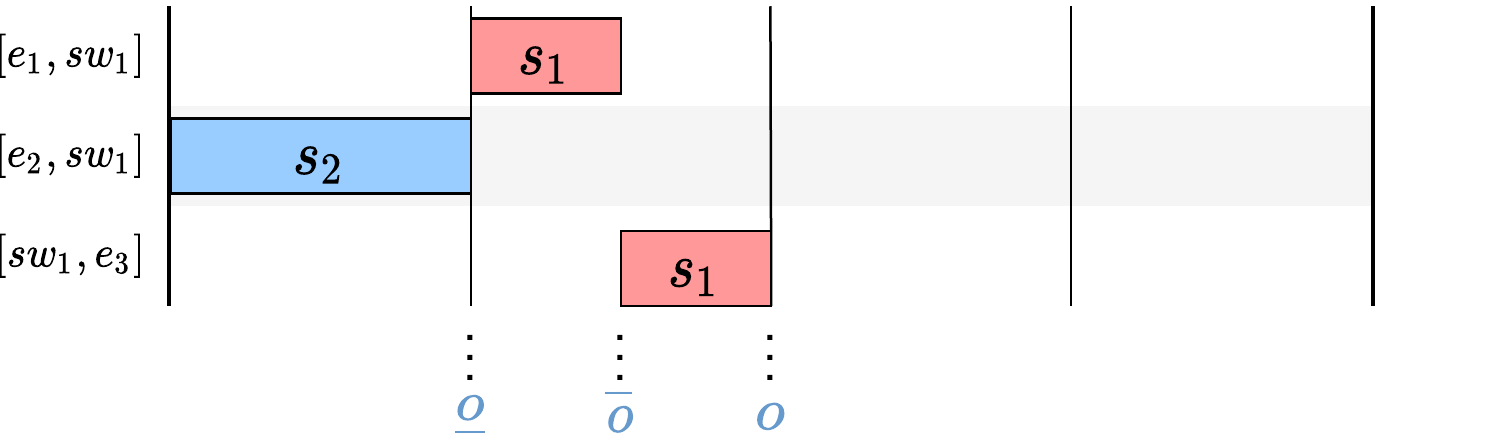}
         \label{fig:backtrack1}
         \caption{Step 1}
     \end{subfigure}
     \hfill
     \begin{subfigure}[b]{0.75\textwidth}
         \centering
         \includegraphics[width=\textwidth]{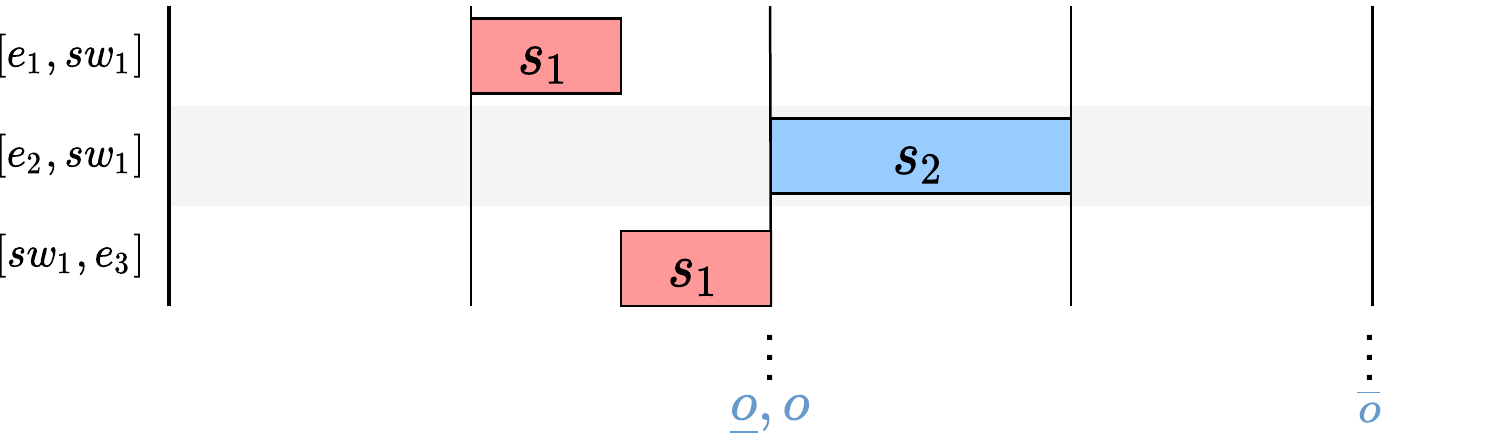}
         \label{fig:backtrack2}
         \caption{Step 2}
     \end{subfigure}
     \hfill
     \begin{subfigure}[b]{0.75\textwidth}
         \centering
         \includegraphics[width=\textwidth]{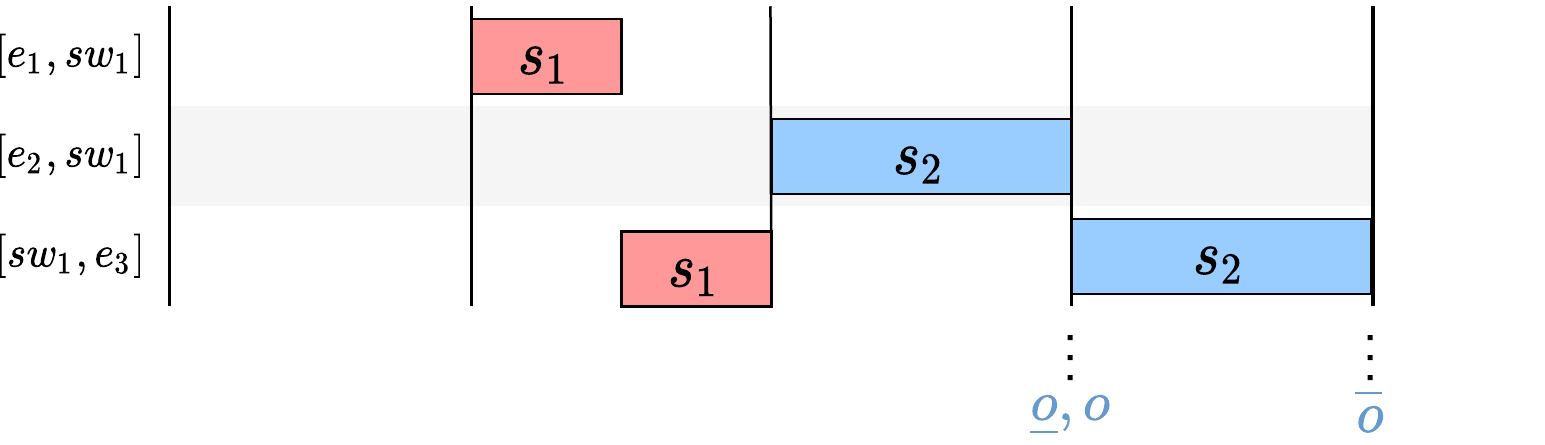}
         \label{fig:backtrack3}
         \caption{Step 3}
     \end{subfigure}
    \caption{Backtrack example: Scheduling $s_2$}
    \label{fig:backtrack}
\end{figure}

\autoref{fig:backtrack} gives an example of this process. In step 1, $s_1$ has already been scheduled, and we are in the process of scheduling $s_2$. We have scheduled the first block on $l_{e_2,sw_1}$ and are now trying to schedule the second one on $l_{sw_1,e_3}$. The lower bound of our offset $\underline{o}$ is set to the end-time of the first block. The upper bound $\overline{o}$ is set to the latest time after which $l_{sw_1,e_3}$ is still available after the offset of the first block, i.e., the start time of $s_1$ on that link. Finally, we find the earliest offset $o$ to be only after the end time of $s_1$. It cannot be earlier since then the blocks of $s_2$ and $s_1$ would overlap. However, scheduling $s_2$ at that time is not possible since it would mean that the two streams interleave at the same port. Consequently, in step 2, we backtrack and reschedule the first block of $s_2$ by setting the lower bound on its offset to the earliest time when its port is available and remains so until $o$. In step 3, we are able to schedule the second block of $s_2$ without problems.

Once we have successfully found an offset for each block, we can update the schedule (\autoref{alg:line:update_schedule}). This will remove the found blocks $B$ from the feasible region.

\subsubsection{Calculating the earliest offset}
\label{sec:earliestoffset}

Calculating the earliest offset (\autoref{algo:offset} shows the function) for a given block is an important part of the heuristic. It takes a block $b$ as an input and calculates the feasible region for that block (line 2). It then returns the lowest possible time that is within the feasible region and greater or equal than the lower bound (lines 3-6).

\begin{algorithm}[h]
\SetAlgoLined
\SetNoFillComment
\SetKwProg{Fn}{Function}{}{}
    \Fn{EarliestOffset(b)}
    {
        \tcc{ordered set of intervals}
        I = GetFeasibleRegion(b)\; 
        \ForEach{$i \in I$}
        {
            $o$ = max($b.\underline{o}$, i.begin)\;
            \uIf{i.contains($o$)}
            {
                \KwRet $o$\;
            }
        }
    }
\caption{ASAP Heuristic - EarliestOffset}
\label{algo:offset}
\end{algorithm}

The function to calculate the feasible regions for a given block $b$ is detailed in \autoref{algo:feasible_region}.
We start by getting all free intervals on the node/link $b.l$ for the period $b.e.T$ of the block (line 3). This ensures that the feasible region does not include any previously scheduled blocks on that node/link. The function then proceeds to fill the data structure $R_{feas}$ with the free intervals, while cutting of a piece with the length of the block $b$ from the end of each such interval (lines 4-7). This makes the feasible region represent all feasible values for the \textit{offset} of the block. 

\begin{figure}
     \centering
     \begin{subfigure}[b]{0.62\textwidth}
         \centering
         \includegraphics[width=\textwidth]{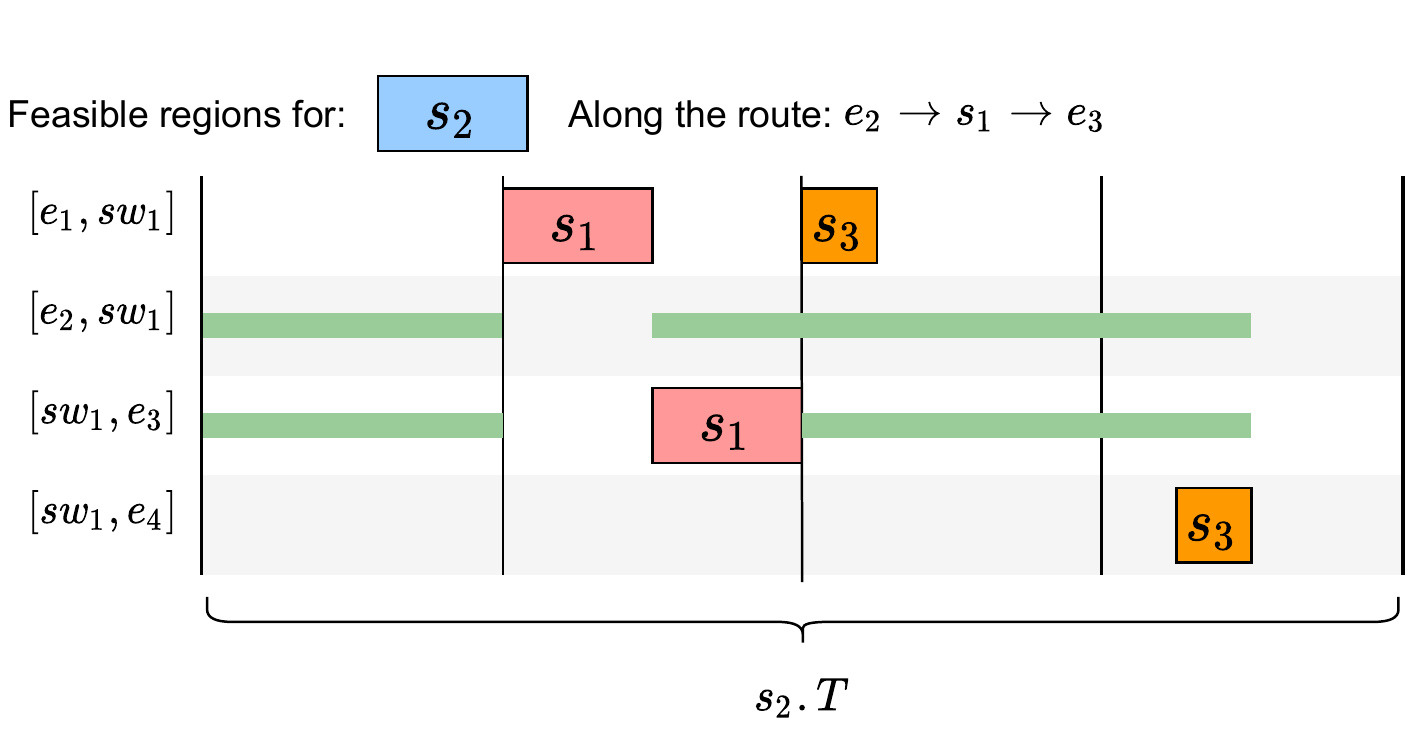}
         \label{fig:feasible_region1}
         \caption{Route 1}
     \end{subfigure}
     \hfill
     \begin{subfigure}[b]{0.62\textwidth}
         \centering
         \includegraphics[width=\textwidth]{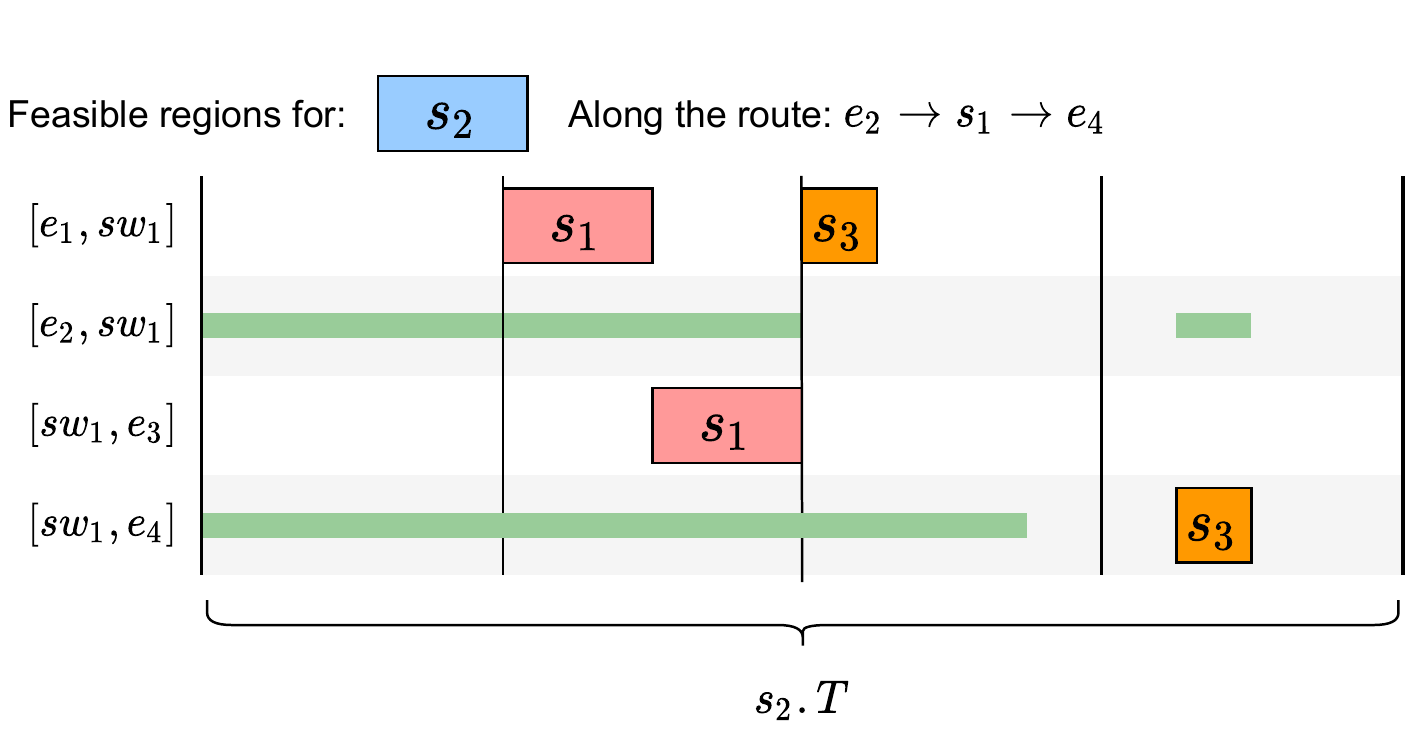}
         \label{fig:feasible_region2}
         \caption{Route 2}
     \end{subfigure}
    \caption{Feasible region example}
    \label{fig:feasible_region}
\end{figure}

\subsection{Optimizing the latency for secure streams}
\label{sec:optimizelatency}

If the block is assigned to a link, we have to cut down the feasible region further. Due to the TSN isolation constraint, it is not allowed to transmit two different streams on the same port at the same time. Thus, we iterate here over all the subsequent blocks $b_{next}$ of the current block $b$, i.e., the blocks on the next links/ES on the route of the stream associated with the block (line 9). If the next block is also assigned to a link (not to an ES), we iterate through all already scheduled blocks $b_{other}$ on that link $b_{next}.l$. These are blocks from other streams with whose predecessors, wherever they are scheduled, we are not allowed to overlap. Thus, we cut the interval ($b_{other}.prev.o,\ b_{other}.o)$ from the feasible region (line 13).

\begin{algorithm}
\SetAlgoLined
\SetNoFillComment
\SetKwProg{Fn}{Function}{}{}
    \Fn{GetFeasibleRegion(b)}
    {
        $R_{feas}$ = $\emptyset$\;
        $B_{free}$ = GetFreeIntervals($b.l$, $b.e.T$)\;
        
        \tcc{(i) Add all free intervals that could contain block $b$}
        \ForEach{$iv \in B_{free}$}
        {
            \uIf{$iv.end - Length(b) \geq iv.begin$}
            {
                $R_{feas}$ = AddToFeasibleRegion($R_{feas}$, (iv.begin, iv.end - Length(b)))\;
            }
        }
        \BlankLine
        \uIf{IsLink(b.l)}
        {
        \tcc{(ii) Cut out the interval blocked by other streams on the next port (TSN Stream Isolation)}
        \ForEach{$b_{next} \in b.next$}
        {
            \uIf{$IsLink(b_{next}.l)$}
            {
                \ForEach{$b_{other} \in GetAllBlocksForLink(b_{next}.l)$}
                {
                    \uIf{$b_{other} \neq b_{next}$}
                    {
                        $R_{feas}$ = CutFromFeasibleRegion($R_{feas}$, ($b_{other}.prev.o, b_{other}.o$))\;
                    }
                }
            }
            
        }
        }
        \KwRet $R_{feas}$\;
    }
\caption{ASAP Heuristic - GetFeasibleRegion}
\label{algo:feasible_region}
\end{algorithm}

\autoref{fig:feasible_region} provides two examples of feasible regions, shown in green, for a stream $s_2$ on two different routes. Looking at \autoref{fig:feasible_region}(a), note the free space at the end of the period and before $s_1$ on $l_{sw_1,e_3}$. Choosing an offset anywhere in this space would result in $s_2$ being scheduled outside its period or overlapping with $s_1$. Choosing an offset in the first free space on $l_{e_2, sw_1}$ would result in $s_1$ and $s_2$ being transmitted to the same port at the same time, breaking the TSN isolation constraint. Note how in \autoref{fig:feasible_region}(b) this is not the case, since $s_2$ is transmitted to a different port ($l_{sw_1, e_4}$) than $s_1$.

\begin{figure}
     \centering
     \begin{subfigure}[b]{0.62\textwidth}
         \centering
         \includegraphics[width=\textwidth]{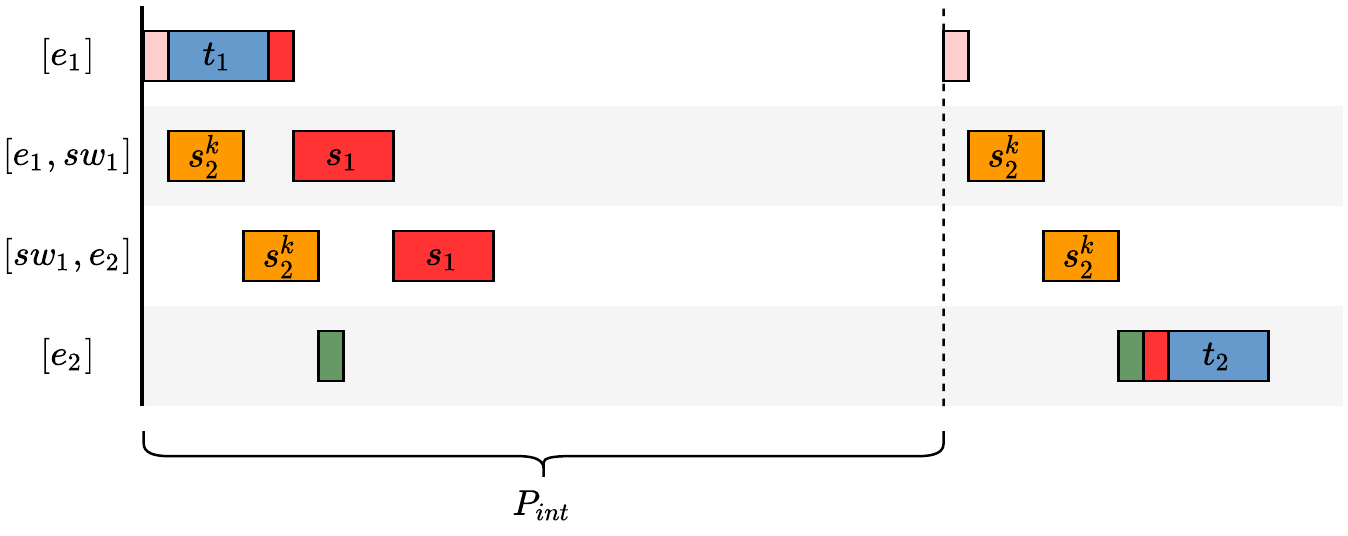}
         \label{fig:latency_unoptimized}
         \caption{Non-optimized stream}
     \end{subfigure}
     \hfill
     \begin{subfigure}[b]{0.62\textwidth}
         \centering
         \includegraphics[width=\textwidth]{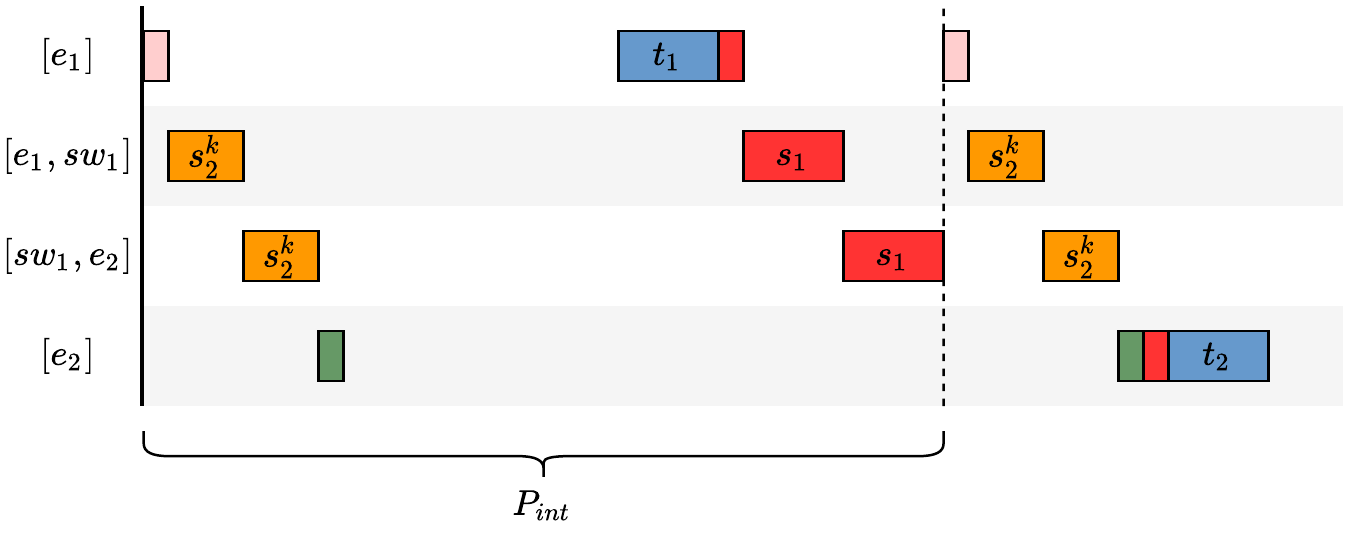}
         \label{fig:latency_optimized}
         \caption{Optimized stream}
     \end{subfigure}
    \caption{Latency optimization for secure streams}
    \label{fig:latency}
\end{figure}

After we have created a new schedule, we can apply some post-processing to minimize its latency. Since TESLA requires a separation of sending and receiving tasks into separate intervals and since we are using an ASAP heuristic, there can be a significant gap between those tasks, as can be seen in \autoref{fig:latency}(a), resulting in an increased latency. To minimize the latency, the algorithm in \autoref{algo:optimize} will go through each secure stream of each application (\autoref{alg:line:isstream}). It will use the \textit{OptimizeLatencyForSecureStream} function in \autoref{algo:optimize_stream} to optimize each stream individually. This function shifts all instances of the given stream as close to the instances on the receiver end-system as possible, without breaking the TESLA constraint. It also has an optional boolean parameter. If that is set, it also shifts the sending task of the given stream (otherwise there would be no latency gain). However, when we are optimizing a redundant stream, i.e. a stream where multiple copies originate at the same task, said task should only be moved when the last copy is optimized (lines \ref{alg:line:rl_greater_1_start}-\ref{alg:line:rl_greater_1_end}). Otherwise, we can shift it immediately (\autoref{alg:line:rl1})

\begin{algorithm}
\SetAlgoLined
\SetNoFillComment
\SetKwProg{Fn}{Function}{}{}
    \Fn{OptimizeLatency($\Sigma, P$)}
    {
        \ForEach{$\lambda in \Lambda$}
        {
            \ForEach{$n \in TopologicalOrder(\lambda, P)$}
            {
                \uIf{IsStream($n.e$)\ \textbf{and}\ $n.e.secure$\ \textbf{and}\ $n.e \in \mathcal{S}^d$} 
                {
                \label{alg:line:isstream}
                    \uIf{$n.e.rl > 1$}
                    {
                        \ForEach{$s_{r} \in \mathcal{S}_{n.e}$}
                        {
                        \label{alg:line:rl_greater_1_start}
                            \uIf{$s_r \neq n.e$}
                            {
                                $n_r$ = GetNode($s_r$, P)\;
                                OptimizeLatencyForStream($n_r$, False)\;
                            }
                        }
                        OptimizeLatencyForStream($n.e$, True)\; \label{alg:line:rl_greater_1_end}
                    }
                    \uElse
                    {
                        OptimizeLatencyForStream($n$, True)\; \label{alg:line:rl1}
                    }
                }
            }
        }
    }
\caption{ASAP Heuristic - OptimizeLatency}
\label{algo:optimize}
\end{algorithm}

The \textit{OptimizeLatencyForSecureStream} function in \autoref{algo:optimize_stream} works internally by looping through the list of receivers of the given stream (\autoref{alg:line:for_each_recv}, multiple in case of a multicast stream). It goes backwards through the linked list of blocks for the stream, starting with the block on the last link before the current receiver (\autoref{alg:line:first_block}). For each block, it will increase the offset as much as possible (move them as far as possible to the right) (\autoref{alg:line:moveright}). After changing the offset we update the schedule (\autoref{alg:line:update}). Then we continue iterating through the linked list (lines \ref{alg:line:normal1}-\ref{alg:line:normal2}). If we arrive at the last block and the $move\_task$ boolean is set, we finish by the offset of the sender task (lines \ref{alg:line:move_task_1}-\ref{alg:line:move_task_2}).

\begin{algorithm}
\SetAlgoLined
\SetNoFillComment
\SetKwProg{Fn}{Function}{}{}
    \Fn{OptimizeLatencyForStream($n$, $move\_task$)}
    {
            \ForEach{$es_{recv} \in receivers(n)$}
            {
                \label{alg:line:for_each_recv}
                b = BlockOnLink($es_{recv}$, n)\;
                $b_{prev}$ = b.prev\; \label{alg:line:first_block}
                $t_{kv}$ = GetKeyVerificationTask($n.src$, $e$)\;
                $b_{kv}$ = GetBlock($t_{kv}$)\;
                i = GetTESLAIntervalForBlock($b_{kv}$)\;
                ub = $i*t_{kv}.T - b_{prev}.L$\;
                \BlankLine
                \While{$b_{prev} \neq \emptyset$}
                {
                    $b_{prev}.o$ = min(ub, $b_{prev}.\overline{o}$)\; \label{alg:line:moveright}
                    UpdateSchedule($b_{prev}$)\;
                    \label{alg:line:update}
                    \uIf{$b_{prev}.prev == \emptyset\ \textbf{and}\ move\_task$\ \textbf{and}\ IsLastReceiver($es_{recv}$)}
                    {
                        \label{alg:line:move_task_1}
                        \tcc{Also move the sender task closer to the first block of the stream}
                        $t_{sender}$ = GetSenderTask(n)\;
                        $b_{sender}$ = GetBlock($t_{sender}$)\;
                        ub = $b_{prev}.o - b_{sender}.L$\;
                        $b_{prev}$ = $t_{sender}$\;
                        \label{alg:line:move_task_2}
                    }
                    \uElseIf{$b_{prev}.prev \neq \emptyset$}
                    {
                        \label{alg:line:normal1}
                        ub = $b_{prev}.o - b_{prev}.prev.L$\;
                        $b_{prev}$ = $b_{prev}.prev$\;
                        \label{alg:line:normal2}
                    }
                }
                
    }    
}
\caption{ASAP Heuristic - OptimizeLatencyForStream}
\label{algo:optimize_stream}
\end{algorithm}

\section{Experimental Results}
\label{sec:Evaluation}
In this section, we evaluate our two solutions to the formulated problem: The Constraint Programming formulation (referred to as \textit{CP}, described in \autoref{sec:Constraint_Formulation}) and the Simulated Annealing metaheuristic (referred to as \textit{SA}, described in \autoref{sec:metaheuristic}). We analyze their scalability, runtime and solution quality and evaluate the impact of added redundancy and security.

Both solutions were implemented in Python 3.9. We developed a software tool with a web-based interactive user interface to display the models and solutions, including a routing graph and the schedule.\footnote{The tool including the obtained results is available on GitHub: \url{https://github.com/nreusch/TSNConf}} For solving the CP formulation we use the CP-SAT solver from Google OR-Tools~\cite{GoogleOR}. For calculating k-shortest-paths in the metaheuristic we use the \textit{shortest\_simple\_paths} function from the NetworkX \cite{NetworkX} library. All evaluations were run on a High Performance Computing (HPC) cluster, with each node configured with 2xIntel Xeon Processor 2660v3 (10 cores, 2.60GHz) and 16 GB memory. Both CP and SA run on one node at a time

\begin{table*}[!ht]
\centering
\begin{tabular}{|c|c|c|c|c|c|c|c|c|}
\hline
\textbf{Test case} & \textbf{Method} & \textbf{\# ES} & \textbf{\# SW} & \textbf{\# Streams} & \textbf{\# Recv. Tasks} & \textbf{\# Tasks} & \textbf{Cost} & \textbf{T} \\ \hline
example           & CP              & 4              & 2              & 6                   & 10                      & 9                 & 467           & 1~s         \\
example           & SA              & 4              & 2              & 6                   & 10                      & 9                 & 477           & 10~m      \\\hline
auto        & CP              & 20             & 32             & 84                  & 102                     & 74                & /             & /          \\
auto        & SA              & 20             & 32             & 84                  & 102                     & 74                & $38\,031$        & 10~m      \\\hline
case\_study       & CP              & 6              & 2              & 29                  & 31                      & 28                & $3\,771$          & 130~s       \\
case\_study       & SA              & 6              & 2              & 29                  & 31                      & 28                & $6\,114$          & 10~m      \\\hline
tiny1             & CP              & 4              & 2              & 2                   & 2                       & 6                 & $1\,708$          & 0.2~s      \\
tiny1             & SA              & 4              & 2              & 2                   & 2                       & 6                 & $1\,708$          & 10~m     \\\hline
tiny2             & CP              & 4              & 2              & 3                   & 4                       & 6                 & $1\,732$          & 0.2 s      \\
tiny2             & SA              & 4              & 2              & 3                   & 4                       & 6                 & $1\,732$          & 10~m    \\\hline
tiny3             & CP              & 4              & 2              & 11                  & 13                      & 15                & $7\,450$          & 14~m     \\
tiny3             & SA              & 4              & 2              & 11                  & 13                      & 15                & $18\,088$         & 10~m     \\\hline
small1            & CP              & 8              & 4              & 10                  & 16                      & 20                & $5\,421$          & 2~s        \\
small1            & SA              & 8              & 4              & 10                  & 16                      & 20                & $13\,303$         & 10~m     \\\hline
small2            & CP              & 8              & 4              & 14                  & 20                      & 23                & $9\,110$          & 60~m     \\
small2            & SA              & 8              & 4              & 14                  & 20                      & 23                & $13\,794$         & 10~m     \\\hline
small3            & CP              & 8              & 4              & 29                  & 48                      & 35                & $7\,705$          & 17.5~m   \\
small3            & SA              & 8              & 4              & 29                  & 48                      & 35                & $13\,781$         & 10~m     \\\hline
medium1           & CP              & 16             & 8              & 23                  & 34                      & 37                & $12\,991$         & 4.5~m    \\
medium1           & SA              & 16             & 8              & 23                  & 34                      & 37                & $22\,883$         & 10~m     \\\hline
medium2           & CP              & 16             & 8              & 30                  & 47                      & 43                & $6\,552$          & 5.2~m    \\
medium2           & SA              & 16             & 8              & 30                  & 47                      & 43                & $19\,455$         & 10~m     \\\hline
medium3           & CP              & 16             & 8              & 36                  & 53                      & 47                & $15\,515$         & 60~m     \\
medium3           & SA              & 16             & 8              & 36                  & 53                      & 47                & $26\,486$         & 10~m    \\\hline
large1            & CP              & 32             & 16             & 47                  & 86                      & 73                & /             & /          \\
large1            & SA              & 32             & 16             & 47                  & 86                      & 73                & $43\,872$         & 10~m     \\\hline
large2            & CP              & 32             & 16             & 33                  & 65                      & 72                & $24\,953$         & 25~m     \\
large2            & SA              & 32             & 16             & 33                  & 65                      & 72                & $41\,026$         & 10~m     \\\hline
large3            & CP              & 32             & 16             & 69                  & 170                     & 104               & /             & /          \\
large3            & SA              & 32             & 16             & 69                  & 170                     & 104               & $34\,860$         & 10~m     \\\hline
huge1             & CP              & 64             & 32             & 84                  & 183                     & 133               & /             & /          \\
huge1             & SA              & 64             & 32             & 84                  & 183                     & 133               & $73\,070$         & 10~m     \\\hline 
huge2             & CP             & 64             & 32             & 99                  & 213                     & 161               & /             & /          \\
huge2             & SA              & 64             & 32             & 99                  & 213                     & 161               & $57\,246$         & 10~m     \\\hline 
huge3             & CP             & 64             & 32             & 99                  & 197                     & 169               & /             & /          \\
huge3             & SA              & 64             & 32             & 99                  & 197                     & 169               & $93\,357$         & 10~m     \\\hline 
giant1            & CP              & 128            & 64             & 144                 & 347                     & 261               & /             & /          \\
giant1            & SA              & 128            & 64             & 144                 & 347                     & 261               & $101\,799$        & 20~m     \\ \hline
\end{tabular}
\caption{Scalability tests}
\label{tab:scalability}
\end{table*}

\subsection{Test cases used for the evaluation}
For the scalability evaluation we used the following test cases, see \autoref{tab:scalability}:
the example presented in \autoref{sec:Problem_Formulation} (\textit{example}), a realistic automotive test case from a large automotive manufacturer (\textit{auto}) \cite{VoicaBahramRedundancy}, a medium-sized automotive case study from~\cite{case_study_tc} (\textit{case\_study}) and 16 synthetic test cases of increasing size and complexity. The topology of the auto test case was adjusted to allow disjunct redundant routes.

For the redundancy/security impact evaluation, we used a further set of 100 synthetic test cases grouped into four batches.

We created the synthetic test cases to be as realistic as possible: They all feature secure streams, redundancy levels between 1 and 3, applications with complex dependencies and a realistic network topology that allows disjunct redundant paths. 

To create realistic topologies, we developed a custom algorithm, as follows. For a given number of switches and end-systems, we create that many random points in 2D space. Then we connect each switch to its closest neighbor until every switch is connected to 4 other switches. Afterwards, we connect each end-system to the closest 3 switches.

To create realistic application DAGs, we used the GGen tool presented in \cite{ggen} and the layer-by-layer method with a depth of 3 and a connection probability of 50\%. If a DAG contains separate subgraphs these are split into separate applications. The application period is chosen randomly among the set \{10, 15, 20, 50ms\}. Nodes of the generated DAG are interpreted as tasks with a random WCET, upper bound at 6\% of the period. Tasks are divided randomly between ES. All outgoing edges of a node in the DAG combined are interpreted as a stream, with the source node as sender task and the destination nodes as receiver tasks. The stream has a random size below or equal to 1.500 Bytes, with a random RL between 1-3 and a 30\% probability to be considered security critical.

We used a global link speed of 1000~Mbit/s. TESLA uses 16~B keys and MACs. A hash computation takes 10~$\mu s$ on every ES.

\subsection{Scalability evaluation}

To evaluate the scalability, we ran both the CP and the SA solutions on the same test cases with the same computing resources. \autoref{tab:scalability} shows the results for each test case for both solutions. The columns \textbf{\# ES, \# SW, \# Streams, \# Tasks} give the total number of ES, SW, streams and tasks respectively. \textbf{\# Receiver Tasks} gives the total sum of stream receiver task (since we consider multicast streams, one stream can have multiple).
The \textbf{Cost} column gives the total cost of the found solution following the cost function in \autoref{alg:cost}. The \textbf{T} column shows the total runtime of the solver.

The CP solution was given a timeout of 60min. If CP failed to find an optimal solution in time, or ran out of memory, we reported Cost and T as empty “/”. The SA solution was given a timeout of 10min (20min for the largest test case, giant1). We used ParamILS \cite{ParamILS}  to optimize the following parameters for the SA heuristic: $T_{start}$, $\alpha$, $k$, $p_{rmv}$ and $w$.
$a$ was set to $50\,000$, $b$ to $10\,000$.

Note that the CP solver will return once the optimal solution is found, while the SA solver will always run until the timeout and return the best feasible (i.e. no missed deadlines or overlap) solution found up to that point. However, SA is able to find a first feasible solution  in a very quick time. For all test cases in \autoref{tab:scalability} it could find one in less than 10~s.

The table shows that CP is able to find solutions up to medium-sized test cases within the given timeout, but it does not scale to the larger test cases. SA is scalable; it is able to find solutions even for the the largest test cases. This scalability comes at an increase in cost by 67\% on average, which can be reduced by giving a longer timeout. This increase is mostly caused by increased application latencies (scheduling cost), which are still within the deadlines, while the routing cost is usually close to or equal to the optimal routing cost from the CP solution. The conclusion is that SA can be successfully used to route and schedule large realistic test cases, and its quality is comparable to the optimal solutions obtained by CP.

\subsection{Impact of adding redundancy and security to a test case}

\begin{table*}[!ht]
\centering
\begin{tabular}{|c|c|c|c|c|c|c|}
\hline
\textbf{} & \textbf{Batch name}                 & \textbf{Security} & \textbf{Redundancy} & \textbf{Cost} & \textbf{Bandwidth} & \textbf{CPU} \\ \hline
0         & batch0 - large streams, small tasks & no                & no                  & 3822.08       & 0.09               & 1.44         \\
1         &                                     & no                & yes                 & +1.34\%       & +25.45\%           & +0.00\%      \\
2         &                                     & yes               & no                  & +256.25\%     & +5.49\%            & +12.99\%     \\
3         &                                     & yes               & yes                 & +258.10\%     & +36.64\%           & +15.93\%     \\ \hline
4         & batch1 - large streams, large tasks & no                & no                  & 17721.32      & 0.08               & 7.06         \\
5         &                                     & no                & yes                 & +-0.00\%      & +15.91\%           & +0.00\%      \\
6         &                                     & yes               & no                  & +64.31\%      & +3.18\%            & +1.39\%      \\
7         &                                     & yes               & yes                 & +64.98\%      & +22.16\%           & +1.75\%      \\ \hline
8         & batch2 - small streams, large tasks & no                & no                  & 18547.6       & 0.01               & 7.51         \\
9         &                                     & no                & yes                 & +0.06\%       & +31.29\%           & +0.00\%      \\
10        &                                     & yes               & no                  & +52.86\%      & +33.92\%           & +0.21\%      \\
11        &                                     & yes               & yes                 & +53.12\%      & +97.96\%           & +0.70\%      \\ \hline
12        & batch3 - small streams, small tasks & no                & no                  & 3713.32       & 0.01               & 1.55         \\
13        &                                     & no                & yes                 & +0.37\%       & +26.62\%           & +0.00\%      \\
14        &                                     & yes               & no                  & +238.77\%     & +27.73\%           & +8.54\%      \\
15        &                                     & yes               & yes                 & +245.32\%     & +85.57\%           & +10.57\%     \\ \hline
\end{tabular}
\caption{Impact of security and redundancy measures}
\label{tab:impact}
\end{table*}

The feasible solutions fulfil the security and redundancy requirements of streams. These requirements introduce extra tasks and streams that need to be routed and scheduled, leading to an overhead compared to the case when we would ignore the security and redundancy requirements of an application. In this set of experiments, we were interested to evaluate the overhead fulfilling the redundancy and security requirements compared to the case these are ignored. These overheads were measured on solution cost, available bandwidth, and CPU resources. Hence, we created four batches of 25 synthetic test cases each. Each test case has a random topology with 8 switches and 16 end-systems, 24 tasks and multiple applications with random DAGs. Streams have a random RL between 1 and 3 and 30\% probability to be considered security critical.

Each batch features either large ($1\,000$-$1\,500$~B) or small (1-250B) streams and either large ($\leq$10\% of period) or small ($\leq$2\% of period) tasks. Each batch was run 4 times using the first feasible SA solution with different combinations of enabled/disabled security and redundancy requirements. Disabled security means that all streams are set to a security level of 0, which disabled redundancy means that all streams are set to a redundancy level of 1.

\autoref{tab:impact} shows the results. We always take the results for the no-security, no-redundancy run as a baseline and note the percentual increase in total cost, total bandwidth occupation percentage and total CPU utilization percentage in the following rows. Bandwidth and CPU utilization are measured as the mean of the total utilization over all links and ESs respectively.

As can be seen, the impact of adding security and redundancy differs significantly, depending on the size of initial streams and tasks. Note that an increase in overhead is expected with an increase in the number and difficulty of the security and redundancy requirements.

Adding redundancy has a negligible impact on cost and CPU utilization but always has a significant impact on bandwidth. Adding security always has a significant impact on cost, as each application with secure streams has to be split into multiple TESLA intervals. The impact of adding security on bandwidth and CPU utilization depends largely on the relative size of streams and WCET of tasks compared to the TESLA overhead. For example, 16 bytes of overhead for a MAC a much more significant for a 100~B stream than for a $1\,000$~B stream.

\subsection{Discussion}
Our proposed SA implementation is able to determine good solutions in a reasonable time, even for large test cases.
In addition, it can find feasible solutions (where all timing, safety and redundancy requirements are satisfied) extremely quickly, within 10 s even for large test cases. This can be useful, e.g., for evaluating several architectures in terms of their monetary costs and redundancy allowed by the physical topology, prototyping or for rapid runtime reconfiguration in case of failures or changes in traffic patterns. Although CP can find optimal solutions, it does not scale for large test cases, and it is not flexible, that is, it will not report solutions which are not feasible. An advantage of SA is its ability to find return near-feasible solutions for those test cases that cannot be solved, i.e.
solutions with some infeasible applications or overlapping streams. SA is able to point out the names of the offending apps/tasks and streams, which can give a good indication on where the configuration has to be improved to become feasible, e.g., by increasing the redundancy in the physical topology or by changing the mapping of tasks to ESs.

\section{Conclusion}
\label{sec:Conclusion}
\noindent In this paper, we addressed the combined TSN routing and scheduling problem for complex applications with redundancy and security requirements. 

We proposed TESLA as an efficient authentication protocol for use-cases with low-powered devices and multicast communication. We proposed a modification to the protocol that makes it more lightweight, made possible by the real-time guarantees of our network.

We developed two methods to solve the combined routing and scheduling problem: A Constraint Programming solution which can solve small and medium-sized test cases optimally and a solution that combines a Simulated Annealing metaheuristic and an ASAP list scheduling which can solve very large test cases.

We formalized the constraints governing our problem and came up with novel ways to handle the complexities introduced by TESLA and redundancy while calculating correct solutions in the heuristic. Furthermore, we developed and shared a useful tool for reuse of our solutions and interactive visualization of routes and schedules.

We evaluated the impact of adding security and redundancy to existing applications and showed that much the overheads depend on the size of existing tasks and streams.

\bibliography{main.bib}{}
\bibliographystyle{acm}

\appendix

\section{Routing constraint formulation for forbidden overlap}
\label{AppendixA}

To achieve a constraint formulation in which overlap is possible do the following:

Replace \eqref{eq:RC2}:
\begin{align*}
    &cost(s_n) = length\_cost(s_n) + 100*overlap\_cost(s_n)\tag{RC2}
\end{align*}

Introduce the following:
\begin{align*}
     &length\_cost(s_n) = \sum_{n \in \mathcal{N} \setminus \{s_n.t_s.e\}}^{} (x(s_n,n) != nil)\tag{RC3}\\
     &overlap\_cost(s_n) = \sum_{n \in \mathcal{N} \setminus \{s_n.t_s.e\}}^{} \sum_{m \in \mathcal{N} \setminus \{n\}}^{}
     link\_cost(s_n, n, m)\tag{RC4}\\
     &link\_cost(s_n, n, m) =
     (xsum(s_n, n, m)-1)*
     (x(s_n,n) == m)\tag{RC5}\\
 \end{align*}
 
Remove \eqref{eq:R6}

\section{Additional functions from metaheuristic formulation}
\label{sec:appendixb}

\subsection{CalculateLowerBound}
\label{sec:lowerbound}
\begin{algorithm}
\SetAlgoLined
\SetNoFillComment

\SetKwProg{Fn}{Function}{}{}
    \Fn{CalculateLowerBound(n, b, P, $\mathcal{R}$)}
    {
        lb = 0\;
        
        \uIf{b.prev == $\emptyset$}
        {
            \tcc{If n is a task or the first stream instance}
            \ForEach{$n_{prev} \in Predecessors(n, P)$}
            {
                b = LastBlock($n_{prev}$)\;
                lb = max(lb, $b.o + Length(b)$)\;
            }
        }
        \uElseIf{IsLink(b.l)}
        {
            \tcc{If $n$ is a stream and $b.l$ is a link}
            \ForEach{$l_{prev} \in PredecessorLinks(b.l,n,\mathcal{R})$}
            {
                $b_{prev}$ = BlockOnLink(b.l, n)\;
                lb = max(lb, $b_{prev}.o + Length(b_{prev})$)\;
            }
        }
        \Else{
            \tcc{If $n$ is a stream and $l$ is a receiver end-system}
            $t_{key}^{verify}$ = GetKeyVerificationTask($n$, $l$)\;
            $b_{key}^{verify}$ = GetBlockForEntry($t_{key}^{verify}$)\;
            i = GetTESLAIntervalForBlock($b_{key}^{verify}$)\;
            lb = $b_{key}^{verify}.o + i*b_{key}^{verify}.e.T + Length(b_{key}^{verify})$\;
        }
        \KwRet max(lb, $b.\underline{o}$)\;
    }
\caption{ASAP Heuristic - CalculateLowerBound}
\label{alg:calc_lb}
\end{algorithm}

\subsection{BlockQueues}
\label{sec:blockqueues}
\begin{algorithm}
\SetAlgoLined
\SetNoFillComment
\SetKwProg{Fn}{Function}{}{}
    \Fn{BlockQueues(n, B, L)}
    {
        \ForEach{$l \in L$}{
        \tcc{Calculate blocks for each frame of n}
        i = 0\;
        \ForEach{$b \in B$}
        {
            offsets[i] = $b.o + i*b.e.T$\;
            endtimes[i] = $b.o + i*b.e.T + Length(b)$\;
            i = i + 1\;
        }
        \BlankLine
        \tcc{Block queues/end-systems}
        \ForEach{$T \in Periods$}
        {
            \For{$i=0$ \KwTo len(offsets)-1}
            {
                o = offset[i]\;
                e = endtimes[i]\;
                
                \uIf{$e\%T < o\%T$}
                {
                \tcc{Handle wrap around period border}
                    CutFromFeasibleRegion(l, $o \% T$, T)\;
                  CutFromFeasibleRegion(l, 0, $e \% T$)\; 
                }
                \Else{
                CutFromFeasibleRegion(l, $o \% T$, $e \% T$)\;
                }
            }
        }
        }
    }
\label{algo:block}
\caption{ASAP Heuristic - BlockQueues}
\end{algorithm}

\end{document}